\input epsf.sty

\def\be{\begin{equation}}
\def\ee{\end{equation}}
\def\bea{\begin{eqnarray}}
\def\eea{\end{eqnarray}}

\newcommand{\h}{$h_{50}^{-1}$}
\def\simlt{\ \raise -2.truept\hbox{\rlap{\hbox{$\sim$}}\raise5.truept   %
\hbox{$<$}\ }}
\def\simgt{\ \raise -2.truept\hbox{\rlap{\hbox{$\sim$}}\raise5.truept   %
\hbox{$>$}\ }}                                                          %

\def\newline{\hfil\break}

\def\la{\mathrel{\hbox{\rlap{\hbox{\lower4pt\hbox{$\sim$}}}\hbox{$<$}}}}
\def\ga{\mathrel{\hbox{\rlap{\hbox{\lower4pt\hbox{$\sim$}}}\hbox{$>$}}}}

\newcommand{\lsim}{{\lower.5ex\hbox{$\; \buildrel < \over \sim \;$}}}
\newcommand{\gsim}{{\lower.5ex\hbox{$\; \buildrel > \over \sim \;$}}}

\def\newline{\hfil\break}
\def\h50{$h_{50}^{-1}$}

\def\om0{$\Omega_0$}

\def\omh{\Omega h}
\def\omh2{\Omega h^2}

\def\simlt{\ \raise -2.truept\hbox{\rlap{\hbox{$\sim$}}\raise5.truept   %
\hbox{$<$}\ }}                                                          %
\def\simgt{\ \raise -2.truept\hbox{\rlap{\hbox{$\sim$}}\raise5.truept   %
\hbox{$>$}\ }}                                                          %

\def\be{\begin{equation}}
\def\ee{\end{equation}}
\def\newline{\hfil\break}
\def\cm-3{cm$^{-3}\,$}

\def\ergcm-3{erg ~cm$^{-3}\,$}
\def\ergcm-2s-1{erg~cm$^{-2}\,$~s$^{-1}\,$}

\def\ergcm2s{\rm{erg}~{\rm cm}^{-2}~ {\rm s}^{-1}}
\def\msun{M_{\odot}}
\def\mug{$\mu {\rm G}$}

\def\farcm{\hbox{$.\mkern-4mu^\prime$}}
\def\es{{\rm 1ES0657-556}~}

\documentclass{aipproc}

\layoutstyle{6x9}

\begin{document}

\title{Dark Matter in Modern Cosmology}


\author{Sergio Colafrancesco}{
  address={ASI-ASDC, Via G.Galilei c/o ESRIN, 00040 Frascati, Italy}
}
%
%

\begin{abstract}
The presence of Dark Matter (DM) is required in the universe
regulated by the standard general relativistic theory of
gravitation. The nature of DM is however still elusive to any
experimental search. We discuss here the process of accumulation
of evidence for the presence of DM in the universe, the
astrophysical probes for the leading DM scenarios that can be
obtained through a multi-frequency analysis of cosmic structures
on large scales, and the strategies related to the multi-messenger
and multi-experiment astrophysical search for the nature of the
DM.
\end{abstract}

\maketitle

\section{Dark Matter in modern cosmology: gained evidence}
\label{sect:intro}

There is overwhelming evidence that we live in a geometrically
flat ($\Omega_0 \approx 1$) universe which is dominated - in the
standard general relativistic cosmological paradigm -  by a dark
form of matter (Dark Matter - DM) and by an obscure form of energy
(Dark Energy - DE).
In such standard cosmological paradigm, DM provides a substantial
fraction $\Omega_{DM} \approx 0.227$ of the overall matter-energy
content (the rest being provided by Dark Energy with $\Omega_{DE}
\approx 0.728$ with the baryonic contribution limited to $\Omega_b
\approx 0.045$, see, e.g., Komatsu et al. 2010) and amounts to
$\sim 83 \%$ of the total mass content of the universe.\\
How did we arrive to such a conclusion?
Is there a general and definite consensus on it?

\subsection{The Dark Matter timeline}

The discovery of the presence of DM can be considered as a typical
scientific revolution (Kuhn, 1970) that induced a change in the
reference (cosmological) paradigm. However, as often occurs in a
paradigm shift, there was no single discovery, but new concepts
were developed and integrated step-by-step \footnote{In this
context, it is necessary, since the beginning of this discussion,
to acknowledge many reviews on the Dark Matter and related
problems, as given by Faber \& Gallagher (1979), Trimble (1987),
Kormendy \& Knapp (1987), Srednicki (1990), Turner (1991), Silk
(1992), Ashman (1992), van den Bergh (2001), Ostriker \&
Steinhardt (2003), Rees (2003), Turner (2003), Colafrancesco
(2006, 2007), Einasto (2004, 2009), among others.}.

To begin, there are actually two dark matter problems, the {\it
local} Dark Matter close to the plane of our Galaxy, and the {\it
global} Dark Matter surrounding galaxies, clusters of galaxies and
large scale structures.

The local Dark Matter (Oort 1932) in the Galactic disk is baryonic
[faint stars or planet (jupiter)-like objects], since a collection
of matter close to the galactic plane is possible if it has formed
by the contraction of pre-stellar matter towards the plane
accompanied by dissipation of the extra energy, so to conserve the
flat shape of the population. It is now clear that the amount of
local Dark Matter is low, and it depends on the mass boundary
between luminous stars and faint invisible stars or planet-like
objects (see Einasto 2009 for a discussion and references
therein).

The global Dark Matter (to which we refer in this paper) is the
dominating mass component in the universe; it is mainly
concentrated in galaxies, clusters and superclusters of galaxies,
and populates all other large-scale structures in the universe.\\
The first evidence for such global DM was obtained by F. Zwicky
(1933). From observations of the radial velocities of eight
galaxies in the Coma cluster, Zwicky found an unexpectedly large
velocity dispersion $\sigma_v = 1019  \pm 360$ km s$^{-1}$ [we
note in passing that Zwicky's velocity dispersion from only eight
galaxies agrees well with the modern value $\sigma_v = 1082$ km
s$^{-1}$, as obtained by, e.g., Colless \& Dunn (1996)]. Zwicky
concluded from these observations that, for a velocity dispersion
of $\sim 1000$ km s$^{-1}$, the mean density of the Coma cluster
would have to be $\sim 400$ times greater than that which is
derived from luminous matter. \footnote{Zwicky actually
overestimated the mass-to-light ratio of the Coma cluster because
he assumed a Hubble parameter $H_0 = 558 km s^{-1} Mpc^{-1}$. His
value for the overdensity of Coma should therefore be reduced from
400 to $\sim$ 50 by using the actual value of the Hubble
parameter. It is interesting to note that Hubble's prestige was so
great at that time, that none of the early authors thought of
reducing Hubble's constant as a way of lowering their
mass-to-light ratios.}
Zwicky concluded, therefore, that in order to hold galaxies
together in the cluster, the Coma cluster must contain huge
amounts of some dark (i.e. not visible), cold matter (Zwicky used
actually the words {\it "Dunkle Kalte Materie"} which might be
regarded as the first reference to Cold Dark Matter ... even
though not in the modern sense).\\
Not until three years later it was found (Smith 1936) that the
Virgo cluster also appears to exhibit an unexpectedly high mass.
Smith made the speculation that the excess mass of Virgo "{\it
represents a great mass of internebular material within the
cluster}" (see van den Bergh 2001).

Six years later than Zwicky's 1933 paper, Babcock (1939) obtained
spectra of the Andromeda galaxy (M31) and found that in its outer
regions the galaxy is rotating with an unexpectedly high velocity,
far above the expected Keplerian velocity. He interpreted this
result either as a high mass-to-light ratio in the periphery or as
a strong dust absorption. Dark Matter in galaxies was also
envisaged.\\
One year later, Oort (1940) studied the rotation and the surface
brightness of the edge-on S0 galaxy NGC 3115. He found that {\it
"The distribution of mass in this system appears to bear almost no
relation to that of light."} He concluded that a value $M/L \sim
250$ should be present in the outer regions of NGC 3115 (note that
this value is reduced by almost an order of magnitude if the
modern distance to this galaxy is adopted). Oort ended his paper
by writing that {\it "There cannot be any doubt that an extension
of the measures of rotation to greater distances from the nucleus
would be of exceptional interest."} However, no connection was
made between the missing mass in this S0 galaxy, and the Zwicky
and Smith missing mass problem in clusters of galaxies.

The DM problem remained as a kind of "anomaly" for roughly a
quarter of century after the initial Zwicky's paper until when
Kahn \& Woltjer (1959) pointed out that M31 and our Galaxy were
moving towards each other, so that they must have completed most
of a (very elongated) orbit around each other in a Hubble time.
These authors found that if M31 and our Galaxy started to move
apart $\sim 15$ Gyr ago, the mass of the Local Group had to be
$\simgt 1.8 \times 10^{12} M_{\odot}$. Assuming that the combined
mass of the Andromeda galaxy and the Milky Way system was $0.5
\times 10^{12} M_{\odot}$, Kahn \& Woltjer (1959) concluded that
most of the mass of the Local Group existed in some invisible
form. They opined that it was most likely that this missing mass
was in the form of hot (with temperature of $\sim 5 \times 10^5$
K) gas. From a historical perspective, it is interesting to note
that also Kahn \& Woltjer did not seem to have been aware of the
earlier papers by Zwicky (1933) and Smith (1936) on the missing
mass problem in clusters of galaxies.

In the 1960's and 1970's, Babcock's optical rotation curve of M31,
and that of Rubin \& Ford (1970), were extended to even larger
radii by Roberts \& Whitehurst (1975) using 21-cm line
observations that reached a radial distance of $\sim 30$ kpc.
These observations clearly showed that the rotation curve of M31
did not exhibit any Keplerian drop-off. From these observations
Roberts \& Whitehurst (1975) concluded that the mass-to-light
ratio had to be $\simgt 200$ in the outermost regions of M31.
Again, it is interesting to note that neither Babcock, nor Roberts
\& Whitehurst, cited the 1933 paper by Zwicky. In other words, no
connection was made between the missing mass in the outer region
of spiral galaxies and the missing mass in galaxy clusters such as
Coma (Zwicky 1933) and Virgo (Smith 1936).
Regarding the flat outer rotation curves of galaxies Roberts
(1999, as quoted in van den Bergh 2001) recalls that this result
{\it "was, at best, received with skepticism in many colloquia and
meeting presentations."}
Nonetheless, the paper by Roberts \& Whitehurst (1975) was
important because it, together with papers on the stability of
galactic disks (e.g. Ostriker \& Peebles 1973), and on the
apparent increase of galaxy mass with increasing radius (e.g.
Ostriker, Peebles \& Yahil 1973), first convinced the majority of
astronomers that a missing mass problem existed.
%

By the mid of 1970's enough research was done for the community to
see that the missing mass ''anomaly'' was not going to go away,
and the majority of astronomers had become convinced that missing
mass existed in cosmologically significant amounts.
{\it ''When... an anomaly comes to seem more than just another
puzzle of normal science, the transition to crisis and to
extraordinary science has begun''} (Kuhn 1970). In other words, it
began to be clear in the mid 1970's that a paradigm shift would be
required to interpret the observations that seemed to support the
ubiquitousness of missing (dark) matter in the universe.

At the beginning of the 1980's the presence of DM was directly
confirmed by many independent sources: the dynamics of galaxies
and of stars within galaxies (Rubin et al. 1978, 1980), the mass
determinations of galaxy clusters based on gravitational lensing
\footnote{It is noted, in passing, that Zwicky (1937) also pointed
out that gravitational lensing might provide useful information on
the total masses of galaxies.} (see, e.g., Bartelmann \& Schneider
2001 for a review and references therein), X-ray studies of
clusters of galaxies (see, e.g., Forman et al. 1985), the
indications of the first N-body simulations of large scale
structure formation and of galaxies and cluster formation (see,
e.g., White et al. 1987).\\
At this time it was also already clear that structures in the
universe form by gravitational clustering, started from initially
small fluctuations of the density of matter (Peebles 1980).
In order to form the presently observed structures, the amplitude
of the density fluctuations must be at least one thousandth of the
density itself at the epoch of recombination, when the universe
started to be transparent.
The relic emission coming from this epoch was first detected in
1965 (Penzias \& Wilson 1965) as an (almost) isotropic cosmic
microwave background (CMB). When finally the fluctuations
(anisotropies) of the CMB were measured by the COBE satellite,
first an large angular scales (Smooth et al. 1992), and by the
Boomerang experiment, also on smaller angular scales (de Bernardis
et al. 2000), they appeared to be two orders of magnitude lower
than expected from the density evolution of the luminous
(baryonic) mass. The solution of the problem was suggested
independently by several theorists.
If we suppose that the dominating mass component of the universe -
Dark Matter - is not made of ordinary matter but of some sort of
non-baryonic matter, then density fluctuations can start to grow
much earlier, and have at the time of recombination the amplitudes
needed to form the structures observed today. The interaction of
this non-baryonic matter with radiation is much weaker than that
of ordinary matter, and radiation pressure does not slow the early
growth of fluctuations.

However, beyond the indubitable success of the DM-dominated
cosmological framework, the fundamental uncertainty on the DM
nature remained.\\
The first suggestion for a non-baryonic DM was made referring to
particles well known at that time to physicists, i.e. neutrinos
($\nu$) (see, e.g., Pontecorvo 1967), for which ''theoretical and
practical considerations'' push progressively some physicists to
agree on a non zero mass neutrino (Quantum mechanics allow then
the oscillation of neutrinos: a $\nu_e$ can, along its travel in
the universe, become a $\nu_{\mu}$ and vice-versa). This can be
considered as the first input from particle physics to the search
for the nature of DM.
Rees (1977) was one of the first astrophysicists to tackle with
the idea of non-baryonic nature for the cosmologically relevant
DM, and suggested that: {\it ''There are other possibilities of
more exotic character - for instance the idea of neutrinos with
small ($\sim$ few eV) rest mass has been taken surprisingly
seriously by some authors''.}
However, this scenario of hot DM soon led to major problems,
because neutrinos move with very high velocities which prevents
the formation of small structures as galaxies (Silk 1968).

Thus some other hypothetical non-baryonic particles were
suggested, such as axions (Abbott \& Sikivie 1983, Preskill, Wise
\& Wilczek, 1983; Dine \& Fischler, 1983) or Weakly Interacting
Massive Particles (e.g. WIMPs, Goldberg 1983, Ellis et al. 1983).
The essential character of these particles to make them good for
cosmology is that they have much lower velocities than neutrinos.
Because of this, the new version of Dark Matter scenario was
called Cold, in contrast to neutrino-dominated Hot Dark Matter
scenario.\\
Numerical simulations of the evolution of the structure of the
universe confirmed the formation of filamentary superclusters and
voids in the Cold Dark Matter scenario of structure formation
(e.g. Ostriker 1993 and references therein). The suggestion of the
Cold Dark Matter has solved most problems of the new cosmological
paradigm.

One unsolved problem remained. Estimates of the total matter
density (ordinary plus DM) yield values $\Omega_m \sim 0.27$ of
the critical density. This value - not far from unity, but
definitely smaller than unity - is neither favoured by theory nor
by the data, including the measurements of the cosmic microwave
background, the galaxy dynamics and the expansion rate of the
universe obtained from the study of distant type Ia supernovae
(SNe). To fill the matter/energy density gap between unity and the
observed matter density it was assumed that some sort of Dark
Energy (DE) exists. \footnote{This assumption was not new: already
Einstein added to his cosmological equations a cosmological
constant term $\Lambda$ (corresponding to a vacuum energy).}
By the early 2000's, refined CMB anisotropy experiments (e.g.,
Boomerang, Maxima, DASI, and then WMAP) demonstrated that the CMB
anisotropy spectrum is the largest detector for the presence of
Dark Matter in their fluctuation spectrum, and by combining their
measurements with independent cosmological distance measurement
using type Ia SNe (Perlmutter 2000), a first determination of the
overall matter-energy composition of the universe was possible
(see Komatsu et al. 2010 for a recent determination of the
cosmological parameter set).
The inclusion of the DE term in the general relativistic
cosmological scenario has filled the last gap in the modern
cosmological paradigm.

\subsection{False alarms and diversionary manoeuvres}

No good detective story is complete without at least one false
clue or some diversionary manoeuvres.

Oort (1960, 1965) believed that he had found some dynamical
evidence for the presence of missing mass in the disk of the
Galaxy. If true, this would have indicated that some of the dark
matter was dissipative in nature. However, late in his life, Jan
Oort confessed (as reported by van den Bergh 2001) that the
existence of missing mass in the Galactic plane was never one of
his most firmly held scientific beliefs. Detailed observations,
that have been reviewed by Tinney (1999), show that brown dwarfs
cannot make a significant contribution to the density of the
Galactic disk near the Sun.

The presence of large amounts of matter of unknown origin has
given rise to speculations on the validity of the Newton law of
gravity at very large distances.
One of such attempts is the Modified Newton Dynamics (MOND) model,
suggested by Milgrom (1983) and Milgrom \& Bekenstein (1987) (see,
e.g., Sanders 1990 for a discussion). Indeed, MOND is able to
explain spiral galaxies quite well imposing a minimum acceleration
scale $a_0 \sim c H_0$, without assuming the presence of some
hidden matter. If the MOND scale $a_0$ is allowed to run, galaxy
clusters might be explained as well. However, there exist several
arguments which make this model unrealistic (see Einasto 2009 for
a discussion). In addition, a full relativistic theory is needed
in any case to construct a consistent cosmology.
A tensor-vector-scalar theory (TeVeS, see Bekenstein 2004) has
been then proposed and it is successful in reproducing MOND in the
proper limit even for photons, and respecting also the classical
tests of general relativity (GR).\\
Whether a consistent cosmology can be constructed with this theory
remains to be seen (in fact, both MOND and TEVES models are not
covariant formalisms of a general relativistic theory of
gravitation, see Capozziello et al. 2008).

Extended Gravity scenarios have been therefore suggested to
explain the large-scale cosmological problem of the accelerated
expansion of the universe and have been also worked out to
describe the morphology and dynamics of very peculiar systems,
like the bullet cluster (Brownstein \& Moffat 2007).
Such a scenario is today the best alternative to a pure particle
DM scenario.

\subsection{The Astro-Particle connection}

All the available information indicates that the standard scenario
for structure formation in the GR framework requires that the
global Cold Dark Matter must be non-baryonic, its density
fluctuations start to grow much earlier than perturbations in the
baryonic matter, and have at the recombination epoch large enough
amplitudes to form all structures seen in the universe.\\
However, the actual nature of the CDM particles still remains
unknown. Physicists have attempted to discover particles which
have the properties needed to explain the structure of the
universe, but so far without success.
This means that a true Astro-Particle connection should be
developed in the search for the nature of DM.
In conclusion, even though the direct information on the dark
components of the universe (DM and DE) comes solely from
astronomical observations, it is clear that a definite
understanding of the nature of DM will come through the discovery
and the multi-messenger study of the fundamental particles of
which DM consists of.

\subsection{A last remark}

The discovery of DM has the general character of a typical
scientific revolution connected with changes of paradigms, as
discussed by Kuhn (1970) in his book "The Structure of Scientific
Revolutions". There are not so many areas in modern astronomy and
cosmology where the development of ideas can be described in these
terms.

\section{Dark Matter in modern cosmology: the present}

\subsection{Motivations}

All the reliable indications for the presence of the dark
components of the universe, and especially of DM, come solely from
astronomical evidence.

The main indications are: i) galaxy rotation curves; ii) dwarf
galaxy mass estimators; iii) galaxy cluster mass estimators; iv)
lensing reconstruction of the gravitational potential of galaxy
clusters and large scale structures; v) the combination of global
geometrical probes of the universe (e.g., CMB) and distance
measurements (e.g., SNe).\\
Supporting evidence comes also form Large Scale Structure
simulations for the leading structure formation scenario.

\subsection{Dark Matter candidates}

There are five basic properties that DM candidates must have:
DM must be dissipationless, collisionless, cold, must behave like
a fluid on galactic scales and above, must behave sufficiently
classically to be confined on galactic scales (see Baltz 2004).
The first three properties do not place any stringent constraint
on the space of possibilities, while the last two place upper and
lower (loose) bounds, respectively, on the mass of the particle.
Such wide space of possibilities allowed theoreticians to propose
many candidates for the DM particles (see Feng 2010).
Among these, the most viable and widely considered candidates for
the DM are, so far, neutralinos (the lightest particles of the
minimal supersymmetric extension of the Standard Model, MSSM, see
e.g. Jungman et al. 1996), sterile neutrinos (the lightest
right-handed neutrino, see e.g. Shaposhnikov 2007) or even other
forms of light DM (see e.g. Boyanovsky et al.2007).\\
In the following, and for the sake of brevity, we will focus
mainly on the astrophysical probes related to some of these DM
candidates, specifically neutralinos and sterile neutrinos. %
We will also focus our discussion, for the sake of conciseness, on
the best astrophysical laboratories for DM indirect search: i)
galaxy clusters (the largest bound containers of DM in the
universe) and ii) dwarf spheroidal galaxies (the cleanest, nearby
and bound DM halos).

\subsection{Dark Matter Probes}

Direct detection of DM particles is the cleanest and most decisive
discriminant (see, e.g., Munoz 2003 for a review). However, it
would be especially interesting if astronomical techniques were to
reveal some of the fundamental particle properties predicted by
fundamental theories.

The dark side of the universe sends us, in fact, signals of the
presence and of the nature of DM that can be recorded using
different astrophysical probes. These probes are of {\it
inference} and {\it physical} character.\\
{\it Inference probes} [i.e., the CMB anisotropy spectrum (see,
e.g., Hu \& Dodelson 2002, Spergel et al. 2003, Komatsu et al.
2010), the dynamics of galaxies (Zwicky 1933), the hydrodynamics
of the hot intra-cluster gas (see Sarazin 1988, Arnaud 2005 for a
review) and the gravitational lensing distortion of background
galaxies by the intervening potential wells of galaxy clusters
(see Bartelmann \& Schneider 1999 for a review and references
therein)] tell us about the presence, the total amount and the
spatial distribution of DM in the large scale structures of the
universe but cannot provide detailed information on the nature of
DM.\\
{\it Physical probes} tell us about the nature and the physical
properties of the DM particles and can be obtained by studying the
astrophysical signals of their annihilation/decay in the
atmospheres of DM-dominated structures (like galaxy cluster and
galaxies). These probes can be recorded over a wide range of
frequencies from radio to gamma-rays and prelude to a full
multi-frequency, multi-experiment and multi-messenger search for
the nature of DM in cosmic structures.

\subsection{A test case: neutralino DM}
 \label{sec:xdm}

\begin{figure}[ht!]
 \epsfysize=8.0cm \hspace{1.0cm} \epsfbox{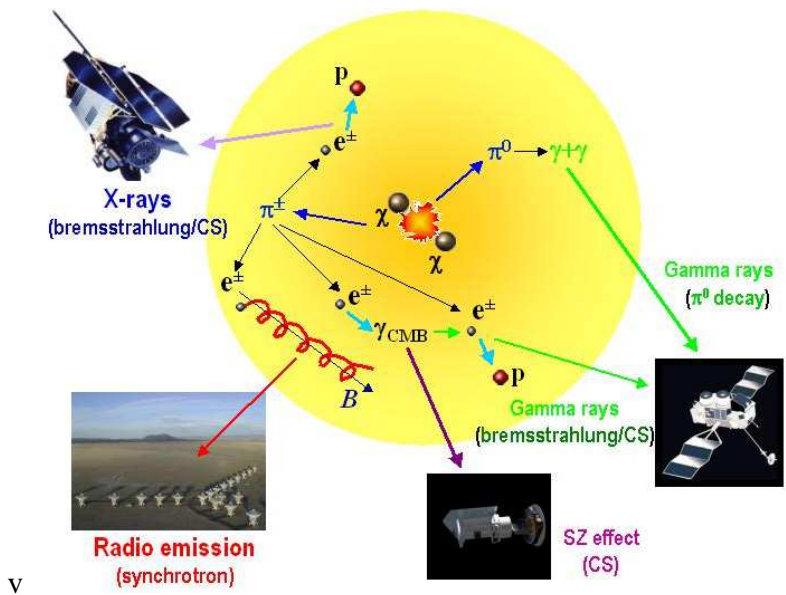}
\caption{A simple model which shows the basic astrophysical
mechanisms underlying the search for the nature of $\chi$ DM
particles through the emission features occurring in large-scale
structures (e.g., galaxy clusters and galaxies). These mechanisms
are, among others: $\gamma$-ray emission from $\pi^0 \to
\gamma+\gamma$, relativistic bremsstrahlung of secondary $e^{\pm}$
and Inverse Compton Scattering (ICS) of CMB photons by secondary
$e^{\pm}$; X-ray/UV emission due to non-thermal bremsstrahlung and
ICS of background photons by secondary $e^{\pm}$; synchrotron
emission by secondary $e^{\pm}$ diffusing in the intra-cluster
magnetic field; Sunyaev-Zel'dovich (SZ$_{DM}$) effect (i.e. ICS of
CMB photons by secondary $e^{\pm}$)}
 \label{fig.dmnature}
\end{figure}

Among the viable competitors for having a cosmologically relevant
DM species, the leading candidate is the lightest particle of the
minimal supersymmetric extension of the Standard Model (MSSM, see
Jungman et al. 1996), plausibly the neutralino $\chi$, with a mass
$M_{\chi}$ in the range between a few GeV to a several hundreds of
GeV (see Baltz 2004 for a review).
Information on the nature and physical properties of the
neutralino DM can be obtained by studying the astrophysical
signals of their interaction/annihilation in the halos of cosmic
structures (galaxy clusters and/or galaxies). These signals
\footnote{Neutralino DM annihilation produces several types of
particle and anti-particle fluxes, whose complete description is
not discussed here for the sake of brevity. We refer the
interested reader to Colafrancesco, Profumo \& Ullio (2006) for
the case of galaxy clusters and Colafrancesco, Profumo \& Ullio
(2007) for the case of dwarf galaxies.} involve, in the case of a
$\chi$ DM, emission of gamma-rays, neutrinos, together with the
synchrotron and bremsstrahlung radiation and the Inverse Compton
Scattering (ICS) of the CMB (and other background) photons by the
secondary electrons produced in the DM annihilation process (see
Fig.\ref{fig.dmnature}).

The $\chi$ annihilation rate is $ R = n_{\chi}(r) \langle \sigma v
\rangle$, where $n_{\chi}(r) = n_{\chi,0} g(r)$ is the neutralino
number density with radial distribution given by the function
$g(r)$, and $\langle \sigma v \rangle$ is the $\chi \chi$
annihilation cross section averaged over a thermal velocity
distribution at freeze-out temperature (Jungman et al. 1996).
The range of neutralino masses and pair annihilation cross
sections in the most general supersymmetric DM setup is extremely
wide. Neutralinos as light as few GeV (see Bottino et al. 2003),
and as heavy as hundreds of TeV (see Profumo 2005) can account for
the observed CDM density through thermal production mechanisms,
and essentially no constraints apply in the case of non-thermally
produced neutralinos.
Turning to the viable range of neutralino pair annihilation cross
sections, coannihilation processes do not allow to set any lower
bound, while on purely theoretical grounds a general upper limit
on $\langle\sigma v\rangle \simlt 10^{-22} (M_{\chi}/ {\rm
TeV})^{-2}{\rm cm}^3/{\rm s}$ can be set (Profumo 2005). The only
general argument which ties the relic abundance of a neutralino
(WIMP) with its pair annihilation cross section is given by
\begin{equation}
\Omega_{DM} h^2\simeq (3\times 10^{-27}{\rm cm}^3/{\rm s})/
\langle\sigma v \rangle
\end{equation}
(see Jungman et al. 1996). This relation can be, however, badly
violated in the general MSSM, or even within minimal setups, such
as the minimal supergravity scenario (see discussion in
Colafrancesco, Profumo \& Ullio 2006).

\subsection{Neutralino annihilations in DM halos}
 \label{sec.chiannih}

Neutralinos which annihilate inside a DM halo produce quarks,
leptons, vector bosons and Higgs bosons, depending on their mass
and physical composition. Electrons and positrons (hereafter
refereed to as electrons for simplicity) are then produced from
the decay of the final heavy fermions and bosons. The different
composition of the $\chi\chi$ annihilation final state will in
general affect the form of the electron spectrum.

Neutral pions produced in $\chi \chi$ annihilation decay rapidly
in $\pi^0 \to \gamma \gamma$ and generate most of the continuum
spectrum at energies $E \simgt 1$ GeV.

Secondary electrons are produced through various prompt generation
mechanisms and by the decay of charged pions, $\pi^{\pm}\to
\mu^{\pm} \nu_{\mu}(\bar{\nu}_{\mu})$, with $\mu^{\pm}\to e^{\pm}
+ \bar{\nu}_{\mu}(\nu_{\mu}) + \nu_e (\bar{\nu}_e)$ and produce
$e^{\pm}$, muons and neutrinos.\\
Secondary electrons are subject to spatial diffusion and energy
losses. Both spatial diffusion and energy losses contribute to
determine the evolution of the source spectrum into the
equilibrium spectrum of these particles, {\em i.e.} the quantity
which is used to determine the multi-frequency spectral energy
distribution (SED) induced by DM annihilation.
The time evolution of the secondary electron spectrum is described
by the transport equation:
\begin{equation}
\frac{\partial n_e}{\partial t} = \nabla \left[ D \nabla
n_e\right] + \frac{\partial}{\partial E} \left[ b_e(E) n_e
\right]+ Q_e(E,r)\;,
 \label{diffeq}
\end{equation}
where $Q_e(E,r)$ is the $e^{\pm}$ source spectrum, $n_e(E,r)$ is
the $e^{\pm}$ equilibrium spectrum and $b_e$ (given here in units
of GeV/s) is the $e^{\pm}$ energy loss per unit time
$
b_e =b_{ICS} + b_{synch} + b_{brem} + b_{Coul} ~,
$
with $b_{ICS} \approx 2.5 \cdot 10^{-17} (E/GeV)^2$, $b_{synch}
\approx 2.54 \cdot 10^{-18} B_{\mu}^2 (E/GeV)^2$, $b_{brem}
\approx 1.51 \cdot 10^{-16} (n_{th}/ cm^{-3})
\left(\log(\Gamma/n_{th})+0.36 \right)$, $b_{Coul} \approx 7\cdot
10^{-16} (n_{th}/ cm^{-3})\left(1+\log(\Gamma/n_{th})/75 \right)$.
Here $n_{th}$ is the ambient gas density and $\Gamma \equiv E/m_e
c^2$.\\
The DM source spectrum, $Q_e(E,r)$, is constant over time, and
under the assumption that the population of high-energy $e^{\pm}$
can be described by a quasi-stationary ($\partial n_e /
\partial t \approx 0$) transport equation, the secondary electron spectrum
$n_e (E,r)$ reaches its equilibrium configuration mainly due to
synchrotron and ICS losses at energies $E \simgt 150$ MeV and to
Coulomb losses at smaller energies.\\
The diffusion coefficient $D$ in eq.(\ref{diffeq}) sets the amount
of spatial diffusion for the secondary electrons: it turns out
that diffusion can be neglected in galaxy clusters while it is
relevant on galactic and sub-galactic scales (see discussion in
Colafrancesco, Profumo \& Ullio 2006, 2007).\\
%
To get a more physical insight on the relevance of spatial
diffusion in large-scale structures, it is useful to consider the
following qualitative solution for the average electron density
 \be
 {d n_e(E,r) \over dE} \approx [Q_e(E,r) \tau_{loss}] \times {V_{s} \over V_s + V_o} \times {\tau_{D} \over \tau_{D}+ \tau_{loss}}
 \label{eq.solution.qualitative}
 \ee
(see Colafrancesco 2005, Colafrancesco et al. 2006) which resumes
the relevant features of the transport equation
(eq.~\ref{diffeq}). Here, $V_s \propto R^3_h$ and $V_o \propto
\lambda^3(E)$ are the volumes occupied by the DM source and the
one occupied by the diffusing electrons which travel a distance
$\lambda(E) \approx [D(E) \cdot \tau_{loss}(E)]^{1/2}$ before
loosing much of their initial energy. The relevant time scales in
eq.~(\ref{diffeq}) are the diffusion time-scale, $\tau_D \approx
R^2_h/ D(E)$, and the energy loss time-scale $\tau_{loss} =
E/b_e(E)$, where we assume  the generic scaling of the diffusion
coefficient $D(E) = \tilde{D}_0 (E/E_0)^{\gamma} B^{- \gamma}$.\\
For $E > E_* = (\tilde{D}_0 E_0/R^2_h b_0
B_{\mu}^{\gamma})^{1/(1-\gamma)}$ [for simplicity we kept leading
terms only, implementing $b(E) \simeq b_0(B_{\mu}) (E /GeV)^2+
b_{Coul}$], the condition $\tau_D > \tau_{loss}$ (and consistently
$\lambda(E) < R_h$) holds, the diffusion is not relevant and the
solution of eq.~(\ref{diffeq}) is $dn_e / dE \sim Q_e(E,r)
\tau_{loss}$ and shows an energy spectrum $\sim Q(E) \cdot
E^{-1}$. This situation ($\lambda(E) < R_h$, $\tau_D >
\tau_{loss}$) applies to the regime of galaxy clusters, i.e.
structures on $\sim$ Mpc scales (see Colafrancesco et al. 2006).\\
For $E < E_*$, the condition $\tau_D < \tau_{loss}$ (and
consistently $\lambda(E) > R_h$) holds, the diffusion is relevant
and the solution of Eq.~(\ref{diffeq}) is $dn_e / dE \sim
[Q_e(E,r) \tau_D] \times (V_{s} / V_o)$ and shows an energy
spectrum $\sim Q(E) \cdot E^{(2-5 \gamma)/2}$ which is flatter or
equal to the previous case for reasonable values $\gamma = 1/3 -
1$. This last situation ($\lambda(E) > R_h$, $\tau_D <
\tau_{loss}$) applies to the regime of dwarf galaxies, i.e.
structures on $\sim$ kpc scales (see Colafrancesco 2005,
Colafrancesco et al. 2007).

Secondary electrons eventually produce radiation by synchrotron in
the magnetized atmosphere of cosmic structures, bremsstrahlung
with ambient protons and ions, and Inverse Compton Scattering
(ICS) of CMB (and other background) photons (and hence an SZ
effect, Colafrancesco 2004). These secondary particles also
produce heating of the ambient gas by Coulomb collisions with the
ambient plasma particles.

\subsubsection{Spectral Energy Distribution from DM annihilation}
\label{sec:multiwave}

\begin{figure}[ht!]
\hbox{
 \epsfysize=7.0cm \hspace{0.0cm} \epsfbox{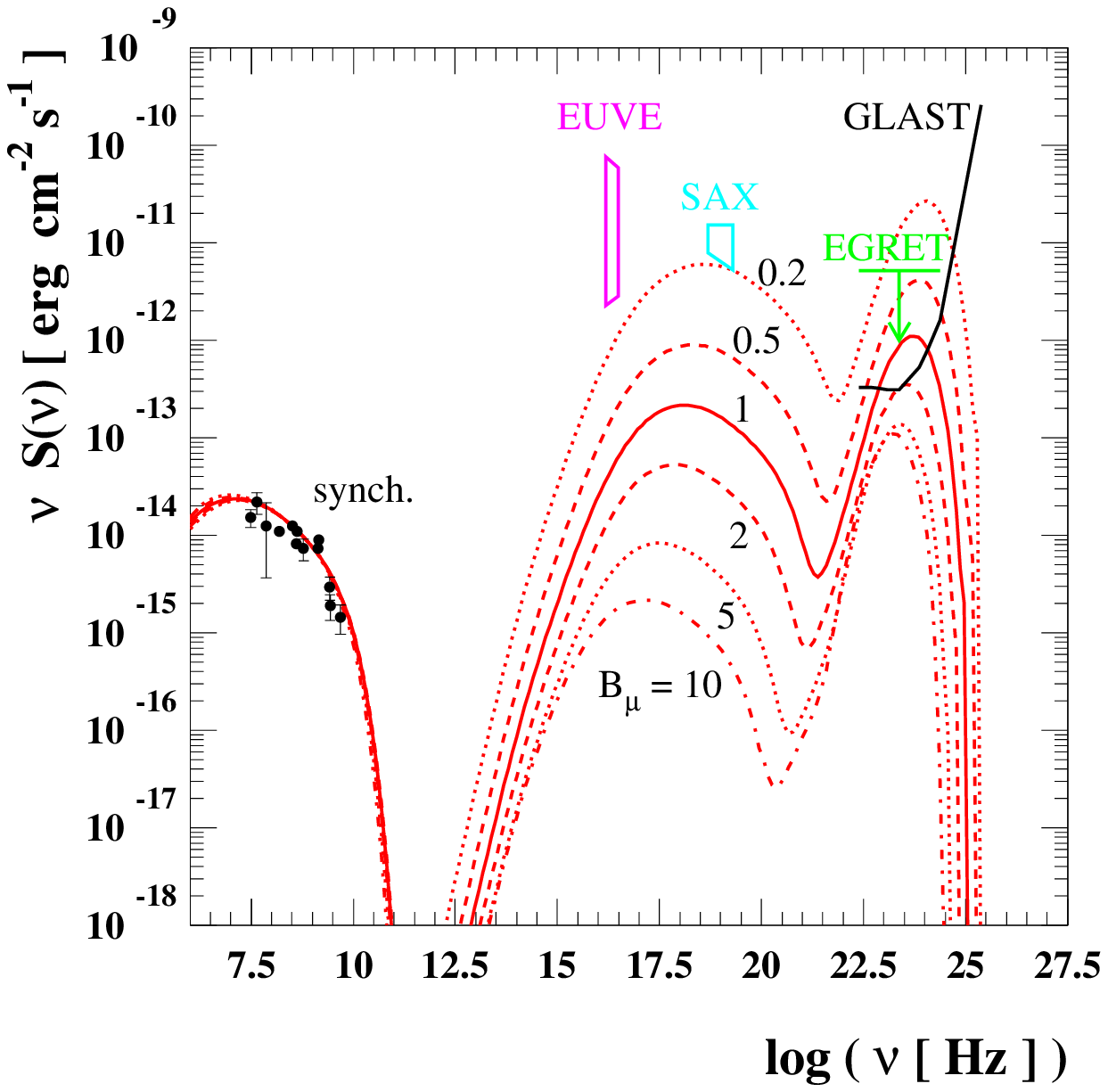}
 \epsfysize=7.0cm \hspace{0.0cm} \epsfbox{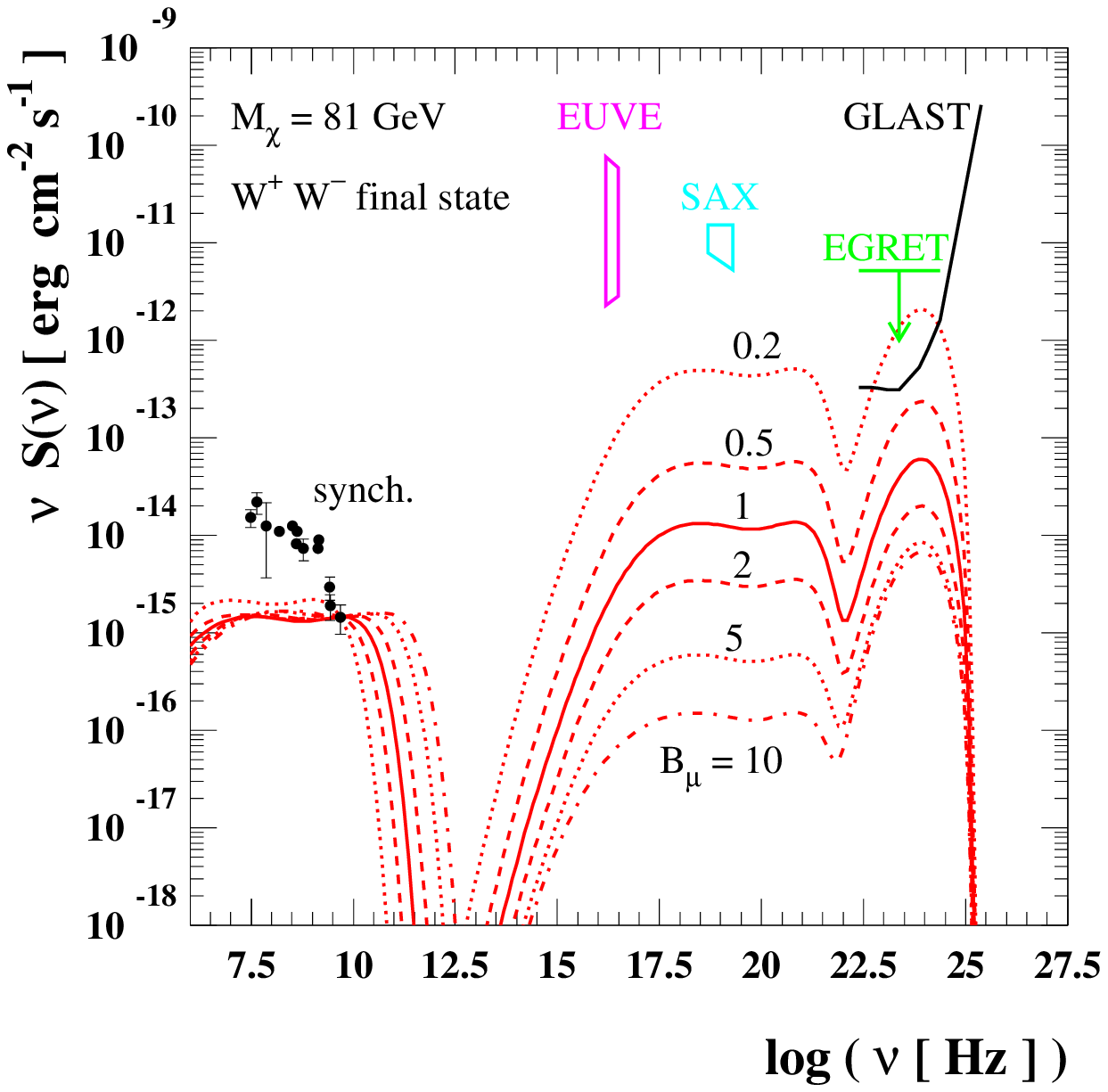}
  }
\caption{\footnotesize{Multi-frequency SED of the Coma cluster for
two representative DM models: $M_{\chi}=40$ GeV ($b \overline{b}$;
left) and $M_{\chi}=81$ GeV ($W^+ W^-$; right). The halo profile
is a NFW profile with $M_{vir} = 0.9 \, 10^{15} \msun h^{-1}$ and
$c_{vir} = 10$, with DM subhalo setup as given in Colafrancesco et
al. (2006). The scaling of the multi-frequency SED with the value
for the mean magnetic field $B_{\mu}$ in Coma is shown for the two
DM models. The neutralino pair annihilation rate has been tuned to
fit the radio halo data  Figures from Colafrancesco et al. (2006).
 }
 }
 \label{fig.gamma_radio_coma}
\end{figure}

\begin{figure}[ht!]
\hbox{
 \epsfysize=7.0cm \hspace{0.0cm} \epsfbox{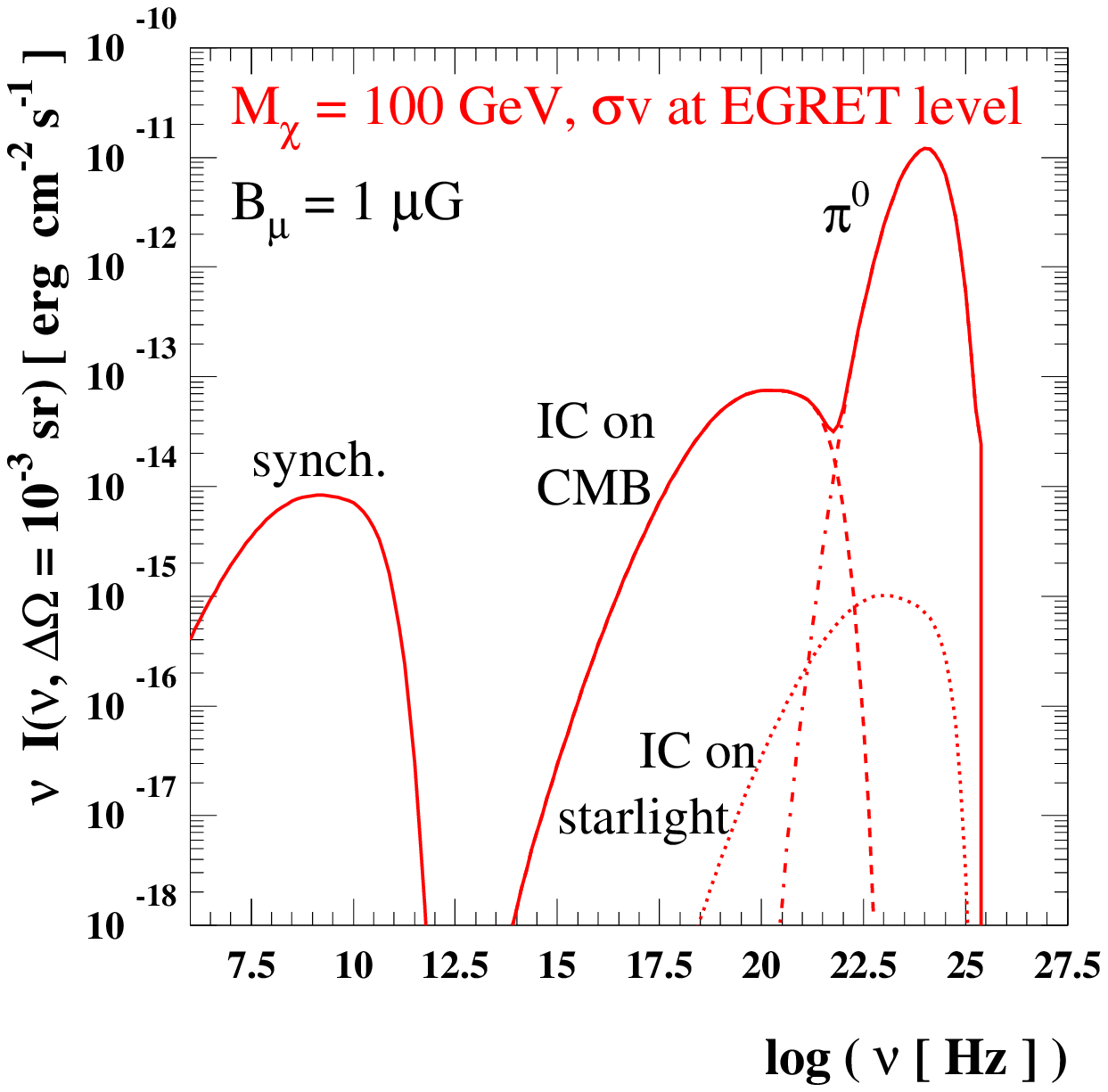}
 \epsfysize=7.0cm \hspace{0.0cm} \epsfbox{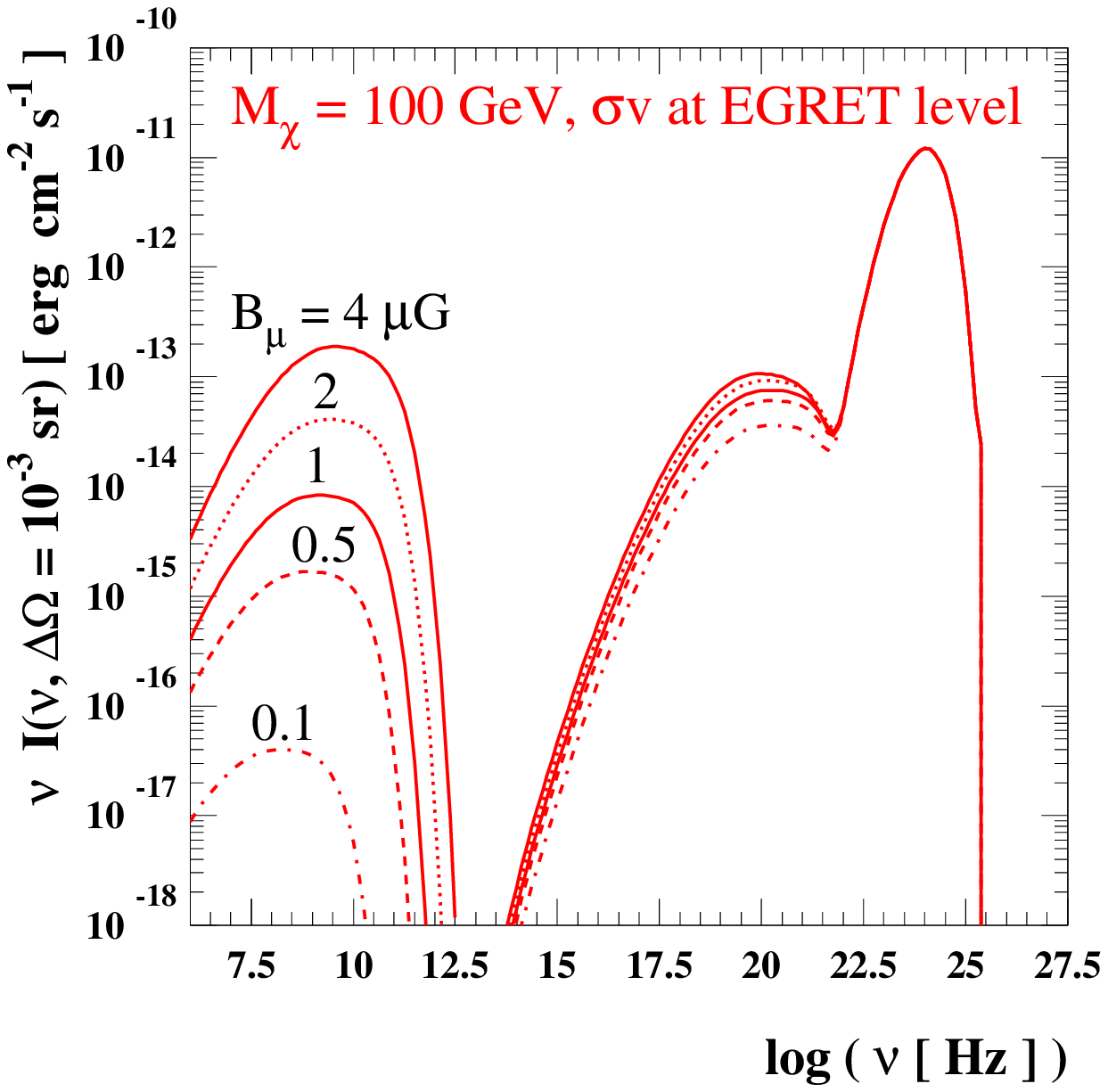}
  }
\caption{\footnotesize{Multi-frequency SED of Draco for a 100 GeV
neutralino annihilating into $b\bar{b}$  and for a magnetic field
of $1 \mu$G (left). The effect of varying the magnetic field
strength is also shown (right). The neutralino pair annihilation
rate has been tuned to give a gamma-ray signal at the level of the
EGRET upper limit. Figures from Colafrancesco et al. (2007).
 }
 }
 \label{fig.gamma_radio_draco}
\end{figure}

The astrophysical signals of neutralino DM annihilation computed
in various DM models can be visible over the entire e.m. spectrum,
from radio to $\gamma$-ray frequencies (see
Figs.\ref{fig.gamma_radio_coma} and \ref{fig.gamma_radio_draco}).
As pointed by Colafrancesco et al. (2006), the relevant physical
properties which determine the features of the emitted radiation
are the composition of the neutralino, its mass, and the value of
the annihilation cross section. We consider here, for the sake of
illustration, a few representative DM models with low (40 GeV),
intermediate (81 GeV) and high (100 GeV) neutralino mass.

\newline{\bf Gamma rays}.
Gamma-ray emission is predominantly due to the hadronization of
the decay products of $\chi \chi$ annihilation with a continuum
$\gamma$-ray spectrum due to the decay $\pi^0 \to \gamma + \gamma$
(see, e.g., Colafrancesco \& Mele 2001, Colafrancesco et al.
2006), even though the direct neutralino annihilation results in a
line emission at an energy $\sim M_{\chi}$.
Gamma-ray emission is also expected from secondary $e^{\pm}$
through bremsstrahlung and ICS of CMB photons up to high energies
(see Figs.\ref{fig.gamma_radio_coma},
\ref{fig.gamma_radio_draco}).\\
For the case of the Coma cluster, the gamma-ray flux produced by
the low mass model ($M_{\chi}=40$ GeV, $b {\bar b}$) is dominated
by the continuum $\pi^0 \to \gamma \gamma$ component and it is a
factor $\sim 5$ lower than the EGRET upper limit of Coma at its
peak frequency once the annihilation cross section is normalized
to fit the radio halo data (see Fig.~\ref{fig.gamma_radio_coma},
left panel). Such gamma-ray flux could be, nonetheless, detectable
by the Fermi (GLAST)--LAT experiment for low values of the
magnetic field $\simlt 0.5$ $\mu$G.
A DM model with intermediate mass ($M_{\chi}=40$ GeV, $W^{\pm}$)
predicts lower gamma-ray flux below the EGRET limit but still
detectable by Fermi (GLAST)-LAT only for very low values of the
magnetic field $\simlt 0.2$ $\mu$G
(Fig~\ref{fig.gamma_radio_coma}, right panel).\\
The rather low neutralino masses of these models make them
difficult to be testable by Cherenkov gamma-ray detectors
operating at high threshold energies.

For the case of smaller cosmic structures, like the Draco dwarf
galaxy, a DM model with $M_{\chi}= 100$ GeV (normalized to the
EGRET upper limit for Draco) predicts that the dominant gamma-ray
emission is still given by the continuum $\pi^0 \to \gamma \gamma$
component while the dominant ICS emission (i.e., ICS on CMB
photons) is a factor $\sim 10^2$ less intense, a fact mainly due
to diffusion effects. The other ICS component considered for Draco
(i.e., ICS on starlight photons) peaks at frequencies comparable
with those of the $\pi^0 \to \gamma \gamma$ component, but is
negligible being a factor $\sim 10^4$ less intense (see
Fig.~\ref{fig.gamma_radio_draco}).\\
Also this model with $M_{\chi} = 100$ GeV is difficult to be
tested by Cherenkov experiments.

\newline{\bf Radio emission}.
Secondary $e^{\pm}$ produced by $\chi \chi$ annihilation can
generate synchrotron emission in the magnetized atmosphere of
galaxy clusters (as well as galaxies) which could be observed at
radio frequencies as a diffuse radio emission (i.e. a radio halo
or haze) centered on the DM halo. Observations of cluster
radio-halos are, in principle, very effective in constraining the
neutralino mass and composition (Colafrancesco \& Mele 2001,
Colafrancesco et al. 2006), under the hypothesis that DM
annihilation provides a major contribution to the radio-halo flux.
Under this hypothesis, a pure energy requirement requires that the
neutralino mass is bound to be $M_{\chi} \geq 23.4 {\rm GeV} (\nu
/ GHz)^{1/2} (B/\mu G)^{-1/2}$ in order that the secondary
$e^{\pm}$ emit at frequencies $\nu \geq 1$ GHz, as observed in
cluster radio halos (see Fig.\ref{fig.gamma_radio_coma}).
\begin{figure}[ht!]
\hbox{
 \epsfysize=7.0cm \hspace{0.0cm} \epsfbox{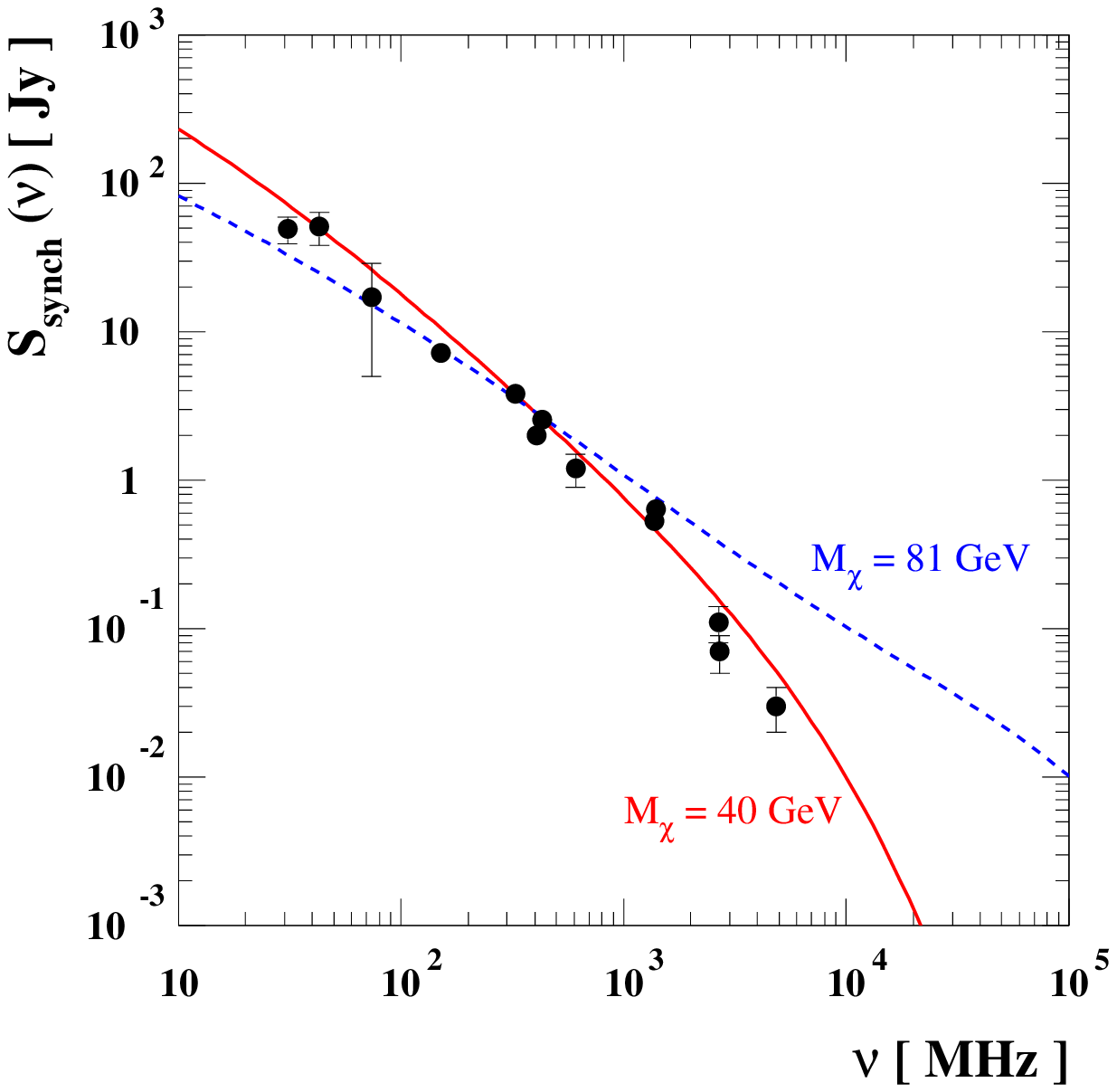}
 \epsfysize=7.0cm \hspace{0.0cm} \epsfbox{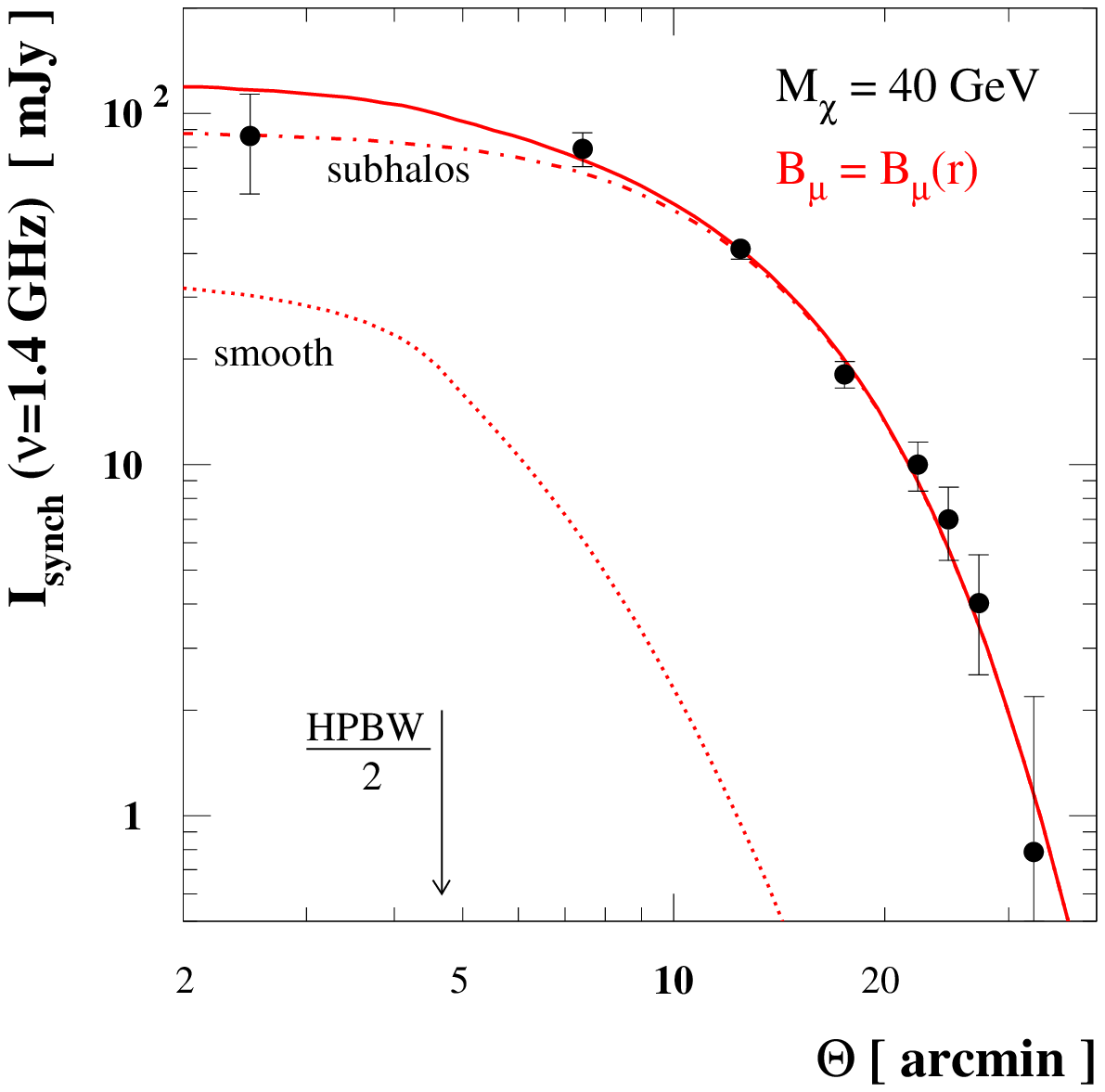}
  }
\caption{\footnotesize{{\bf Left}. The radio flux density spectrum
of Coma in two DM models: soft spectrum due to a $b \bar{b}$
annihilation final state (solid line) with $M_{\chi} = 40$~GeV)
and hard spectrum due to a $W^+ W^-$ channel (dashed line) with
$M_{\chi} = 81$~GeV.
{\bf Right}. Surface brightness distribution of Coma at $\nu =
1.4$~GHz, within a beam equal to $9\farcm35$ (HPBW), for the the
soft neutralino model (solid). We consider a magnetic field $ B(r)
= B_0 (1 + (r/ r_{c1})^2) \cdot [ 1 + ( r /r_{c2})^2]^{-\beta}$
with $B_0 = 0.55$~\mug, $\beta = 2.7$, $r_{c1}=3^\prime$
$r_{c2}=17\farcm5$. The contributions to the radio brightness from
the smooth DM halo component only (dotted) and from subhalos only
(dashed) are also displayed (see Colafrancesco et al. 2006 for
details).
 }
 }
 \label{fig.radio_coma}
\end{figure}
A soft DM model ($b {\bar b}$ with $M_{\chi}=40$ GeV) is able to
reproduce both the overall radio-halo spectrum of Coma and the
spatial distribution of its surface brightness (see
Fig.\ref{fig.radio_coma}), while a hard DM model ($W^+ W^-$ with
$M_{\chi}=81$ GeV) is excluded being its radio spectrum too flat
to reproduce the Coma data.

\begin{figure}[ht!]
\hbox{
 \epsfysize=7.0cm \hspace{0.0cm} \epsfbox{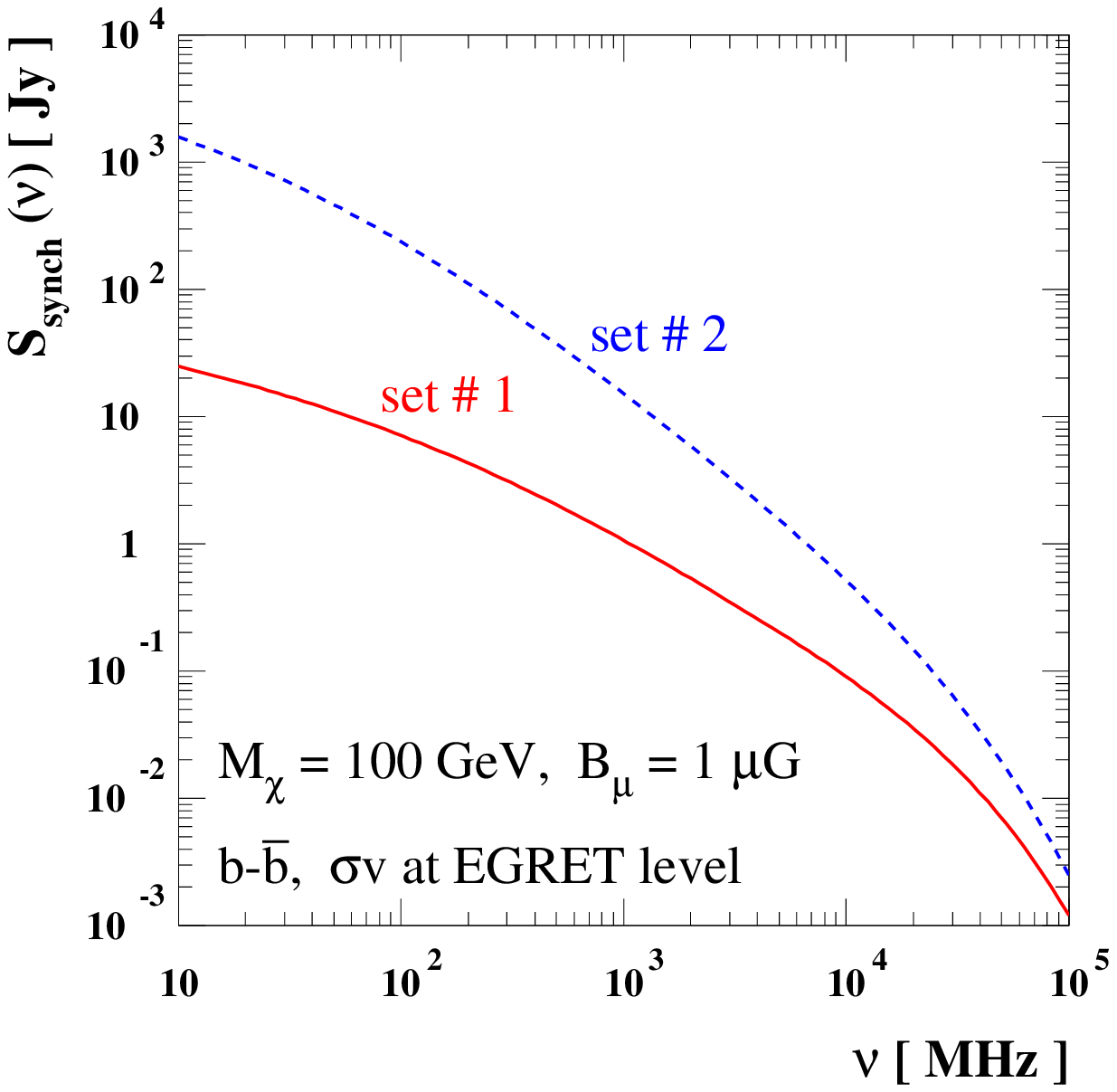}
 \epsfysize=7.0cm \hspace{0.0cm} \epsfbox{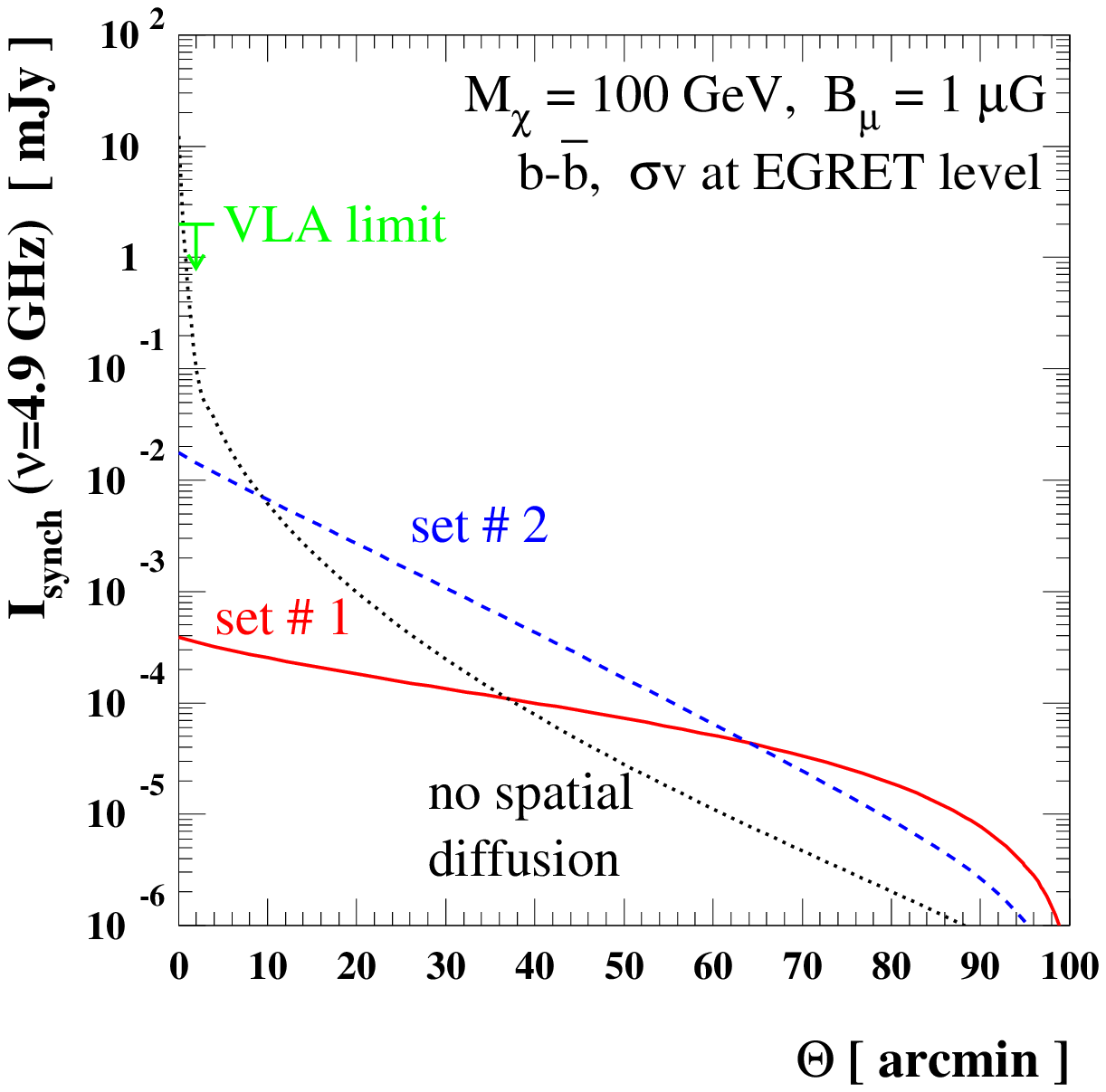}
  }
\caption{Radio flux density spectrum (left) and surface brightness
distribution (right) of Draco at $\nu = 4.9$~GHz for a sample
$\chi$ model with $M_{\chi}= 100$~GeV annihilating into
$b\,\bar{b}$ with a cross section tuned as to give 2 events in the
EGRET experiment. Results are shown for the two choices of
propagation parameters, as labelled; the case when spatial
diffusion is neglected is also shown for comparison in the right
panel (dotted curve). Surface brightness radial profiles are
plotted for a 3~arcsec angular acceptance, corresponding to the
VLA resolution used to search for a point-like radio source at the
center of Draco (the VLA upper limit is also plotted in the
figure).
 }
 \label{fig:radio_draco}
\end{figure}

For the case of dwarf galaxies (e.g. Draco), radio emission is
strongly affected by propagation effects.
\footnote{For the diffusion coefficient we consider here the case
of a Kolmogorov form $D(E) = D_0/B_\mu^{1/3}
\left(E/{\rm{1\;GeV}}\right)^{1/3}$, with $D_0 = 3 \cdot
10^{28}$~cm$^{2}$~s$^{-1}$, in analogy with its value for the
Milky Way (here $B_\mu$ is the magnetic field in units of $\mu$G).
The dimension of the diffusion zone is, consistently with the
Milky Way picture, about twice the radial size of the luminous
component, i.e. $\approx$ 102~arcmin for Draco (set \#1 of
propagation parameters). An extreme diffusion model in which the
diffusion coefficient is decreased by two orders of magnitudes
down to $D_0 = 3 \cdot 10^{26}$~cm$^{2}$~s$^{-1}$ (implying a much
smaller scale of uniformity for the magnetic field), and with a
steeper scaling in energy, $D(E) = D_0
\left(E/{\rm{1\;GeV}}\right)^{-0.6}$ (this is the form sometime
assumed for the Milky Way) is considered for comparison (we label
this propagation parameter configuration set \#2)}
Fig.\ref{fig:radio_draco} shows that for a propagation set up (set
up \#1) with a Kolmogorov spectrum ($D \propto E^{1/3}
B_{\mu}^{-1/3}$) there is a depletion of the electron populations
with a significant fraction leaving the diffusion region, while
for a propagation set up (set up \#2) with a steeper spectrum ($D
\propto E^{0.6} B_{\mu}^{-0.6}$) they are more efficiently
confined within the diffusion region but still significantly
misplaced with respect to the emission region.
As a consequence, also the spectral shape of the radio flux of
Draco is affected by diffusion effects which produce a steeper
spectral slope when the electron populations are more efficiently
confined within the diffusion region  (i.e. propagation set up
\#2) with respect to the case (i.e. propagation set up \#1) where
there is a depletion of the electron populations with a
significant fraction leaving the diffusion region (see
Colafrancesco et al. 2007 for details).

\newline{\bf ICS of CMB: from microwaves to gamma-rays}
Secondary $e^{\pm}$ produced by neutralino annihilation up-scatter
CMB (and other background) photons that will redistribute over a
wide frequency range up to gamma-ray frequencies.
A soft neutralino model with $M_{\chi} = 40$ GeV and $\langle
\sigma v \rangle = 4.7 \cdot 10^{-25} cm^3 s^{-1}$, with
$B_{\mu}=1.2$ yields EUV and HXR fluxes which are more than one
order of magnitude fainter than the Coma data (see
Fig.\ref{fig.gamma_radio_coma}).
Increasing $M_{\chi}$ does not provide a decent fit of the
radio-halo spectrum (see Fig.~\ref{fig.gamma_radio_coma}, right
panel) and yields, in addition, extremely faint EUV, HXR and
gamma-ray fluxes, which turn out to be undetectable even by Fermi
(GLAST) and/or by the next coming high-energy experiments.
Increasing $\langle \sigma v \rangle$ by a factor $\sim 10^2$
(i.e., up to values $\langle \sigma v \rangle \approx 7 \cdot
10^{-23} cm^3 s^{-1}$) can fit the EUV and HXR data on Coma but
the relative $\pi^0 \to \gamma \gamma$ gamma-ray flux exceeds the
EGRET upper limit on Coma.\\
Lowering the magnetic fields down to values $\sim 0.15 \mu$G can
fit both the HXR and the EUV fluxes of Coma but also in this case
the $\pi^0 \to \gamma \gamma$ gamma-ray flux predicted by the same
model exceeds the EGRET limit on Coma, rendering untenable also
this alternative.
Actually, the EGRET upper limit on Coma set a strong constraint on
the combination of values $B$ and $\langle \sigma v \rangle$ so
that magnetic field larger than $\simgt 0.3 $ $\mu$G are required
for the parameter setup of the $b {\bar b}$ model with
$M_{\chi}=40$ GeV (Fig.22 of Colafrancesco et al. (2006) shows the
upper limits on the value of $\langle \sigma v \rangle$ as a
function of the assumed value of the mean magnetic field of
Coma).\\
According to these results, it is impossible to fit all the
available data on Coma for a consistent choice of the DM model and
of the cluster magnetic field (see also discussion in
Colafrancesco et al. 2010).

For the Draco dwarf galaxy the dominant ICS on CMB component
produces fluxes of X-rays at the level of $\sim 10^{-15}-10^{-14}$
erg cm$^{-2}$ s$^{-1}$ when the gamma-ray flux is normalized to
the EGRET upper limit. The constraints obtainable by Fermi (GLAST)
observation of Draco will hence set much more realistic
expectations for the diffuse X-ray emission produced from DM
annihilation in Draco which could eventually be tested with high
sensitivity X-ray, and especially HXR and soft $\gamma$-ray
observations where this ICS spectrum peaks at $E \sim 100$ keV.

\newline{\bf ICS of CMB: SZ effect from DM annihilation}.
Secondary e$^{\pm}$ produced by neutralino annihilation interact
with the CMB photons and up-scatter them to higher frequencies
producing a peculiar SZ effect (as originally realized by
Colafrancesco 2004) with specific spectral and spatial features.
The DM induced spectral distortion writes as
 \begin{equation}
\Delta I_{\rm DM}(x)=2\frac{(k_{\rm B} T_0)^3}{(hc)^2}y_{\rm DM}
~\tilde{g}(x) ~,
\end{equation}
where $T_0$ is the CMB temperature and the Comptonization
parameter is
\begin{equation}
y_{\rm DM}=\frac{\sigma_T}{m_{\rm e} c^2}\int P_{\rm DM} d\ell ~,
\end{equation}
in terms of the pressure $P_{\rm DM}$ contributed by the secondary
$e^{\pm}$.
The function $\tilde{g}(x)$, with $x \equiv h \nu / k_{\rm B}
T_0$, for the DM produced secondary $e^{\pm}$ can be written as
\begin{equation}
 \label{gnontermesatta}
 \tilde{g}(x)=\frac{m_{\rm e} c^2}{\langle
 \varepsilon \rangle} \left\{ \frac{1}{\tau_{DM}}
 \left[\int_{-\infty}^{+\infty} i_0(xe^{-s}) P(s) ds- i_0(x)\right]
 \right\}
\end{equation}
in terms of the secondary $e^{\pm}$ optical depth $\tau_{DM} =
\sigma_T \int d \ell n_e$ (with $n_e$ given by the solution of
eq.\ref{diffeq}), of the photon redistribution function $P(s)=
\int dp f_{\rm e}(p) P_{\rm s}(s;p)$ with $s = \ln(\nu'/\nu)$, in
terms of the photon frequency increase factor $\nu' / \nu$, of the
$e^{\pm}$ momentum ($p$) distribution $f_{\rm e}(p)$, of $i_0(x) =
2 (k_{\rm B} T_0)^3 / (h c)^2 \cdot x^3/(e^x -1)$, and of the
quantity
\begin{equation}
 \langle \varepsilon \rangle \equiv  \frac{\sigma_{\rm T}}{\tau_{DM}}\int P d\ell
= \frac{\int P d\ell}{\int n_{\rm e} d\ell}
 = \int_0^\infty dp f_{\rm e}(p) \frac{1}{3} p v(p) m_{\rm e} c
 \label{temp.media}
\end{equation}
which is the average energy of the secondary electron population
(see Colafrancesco 2004).
\begin{figure}[!t]
\hbox{
 \epsfysize=5.5cm \hspace{0.0cm} \epsfbox{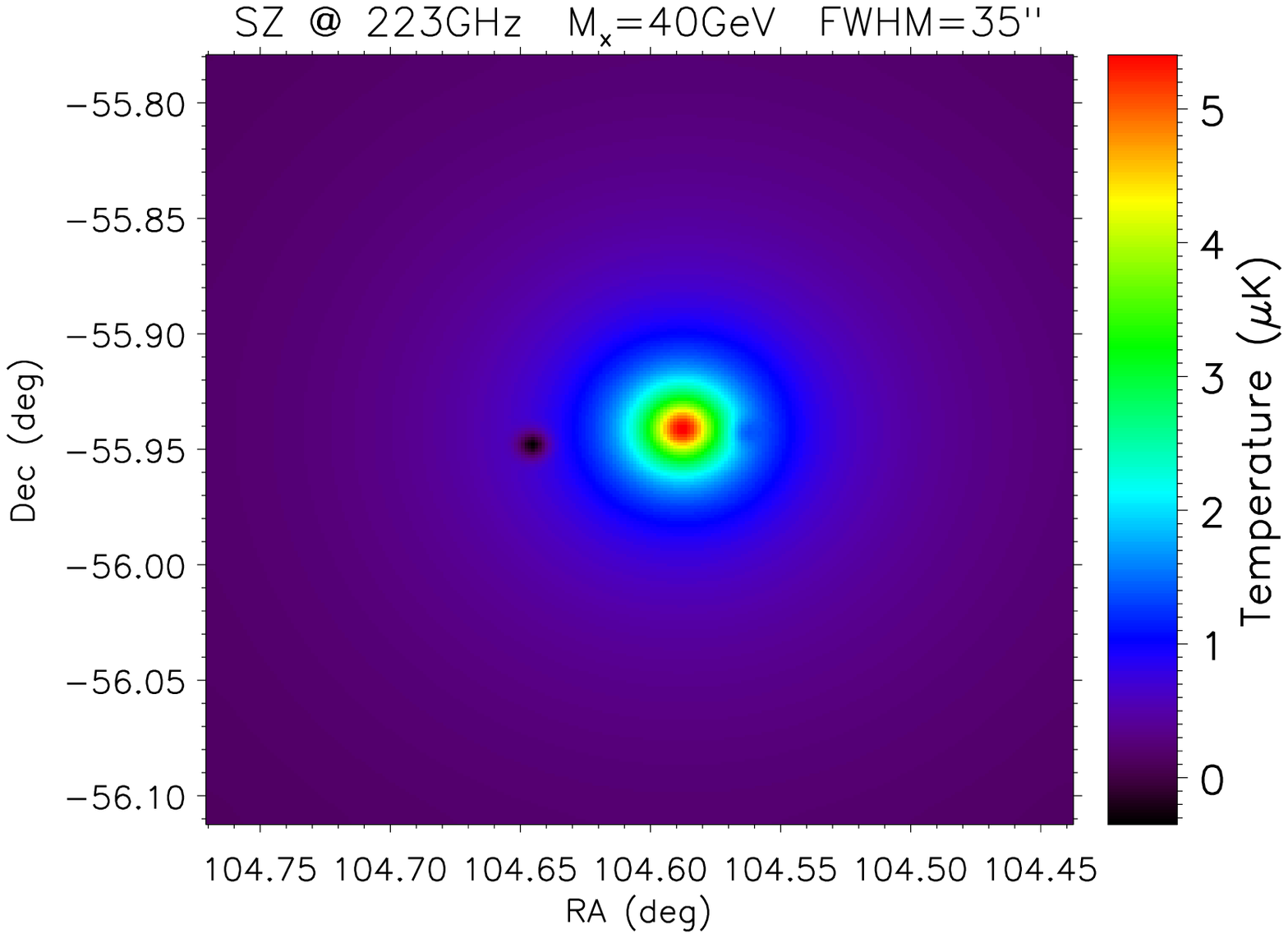}
 \epsfysize=5.5cm \hspace{0.0cm} \epsfbox{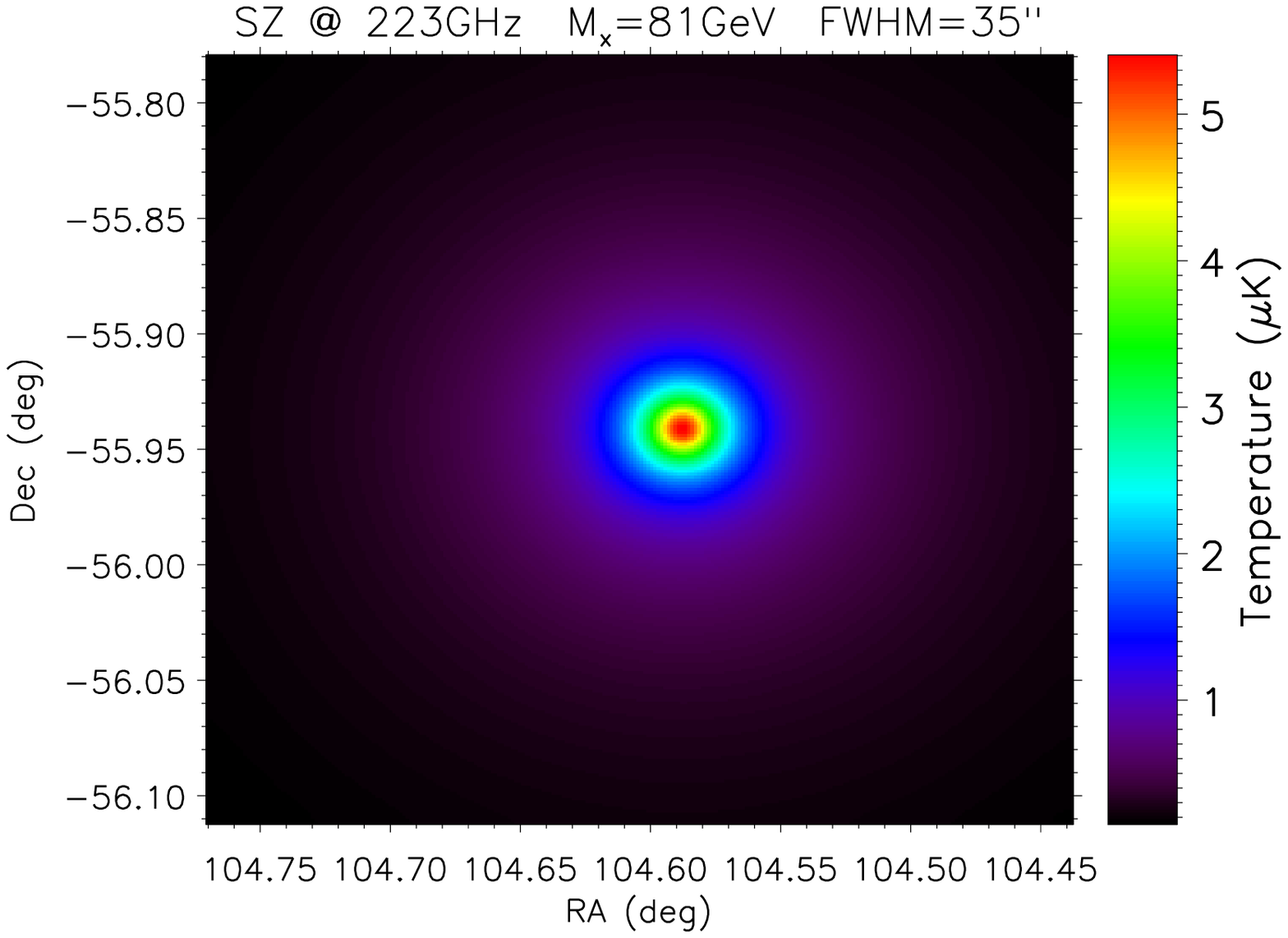}
}
\caption{The simulated SZ maps of the cluster \es observable at
$\nu = 223$ GHz with a telescope of angular resolution similar to
SPT (see Colafrancesco et al. 2007 for details) for a neutralino
mass $M_{\chi}=40$ GeV, $b {\bar b}$ (left panel) and
$M_{\chi}=81$ GeV $W^{\pm}$ (right panel) with the DM density
model setup of Colafrancesco et al. (2006).
 }
 \label{fig:sz_dm}
\end{figure}
Fig.\ref{fig:sz_dm} shows the maps of the CMB temperature change,
 \be
{\Delta T \over T_{0}} = {(e^x-1)^2 \over x^4 e^x} {\Delta I \over
I_0} \,,
 \ee
at 223 GHz as produced by the SZ$_{DM}$ effect in the neutralino
models with mass of 40 GeV ($b {\bar b}$) and 81 GeV ($W^{\pm}$),
compared to the temperature change due to the residual thermal SZ
effect produced by the intracluster gas of the bullet cluster.
The spatial separation between the DM and thermal SZE of this
cluster is due to the fact that the SZ$_{DM}$ effect has a very
different spectral shape with respect to the thermal SZ effect: it
yields a temperature decrement at all the microwave frequencies
$\simlt 600$ GHz, where the thermal SZ effect is predominantly
observed, and produces a temperature increase only at very high
frequencies $> 600$ GHz. This behavior is produced by the large
frequency shift of CMB photons induced by the ICS off the
relativistic secondary electrons generated by the neutralino
annihilation. As a consequence, the zero of the SZ$_{DM}$ effect
is effectively removed from the microwave range and shifted to a
quite high frequency $\sim 600$ GHz with respect to the zero of
the thermal SZ effect, a result which allows, in principle, to
estimate directly the pressure of the two electron populations and
hence to derive constraints on the neutralino DM model (see
Colafrancesco 2004).\\
It is, however, necessary to stress that in such frequency range
there are other possible contributions to the overall SZ effect,
like the kinematic SZ effect and other non-thermal SZ effects
which could provide additional biases.\\
We stress that a SZ$_{DM}$ effect is expected in every neutralino
DM halo and its amplitude depends basically on the optical depth
of the DM-produced secondary electrons, and hence on the detailed
distribution of the equilibrium spectrum of the secondary
electrons.\\
The SZ$_{DM}$ effect produced in a dwarf galaxy, like Draco (see,
e.g., Colafrancesco 2004, Culverhouse, Ewans and Colafrancesco
2006), is however quite low when the spatial diffusion of
secondary electrons is efficiently operating: we find, in fact,
that the SZ signal towards the center of Draco is negligible even
when we normalize the gamma-ray signal at the level of the EGRET
upper limit.

\newline{\bf Heating}.
Low energy secondary electrons produced by neutralino annihilation
heat the intra-cluster gas by Coulomb collisions since the Coulomb
loss term dominates the energy losses at $E \simlt 150$ MeV.
The DM-induced heating rate at the center of galaxy clusters is
usually higher than the intra-cluster gas cooling rate for steep
DM density profiles (like the NFW one). The radius of the region
in which such steep DM density profile produce an excess heating
increases with increasing neutralino mass, once the DM-produced
SED are normalized to fit the radio halo data.
The heating effect provides, hence, strong constraint to the
annihilation cross-section for neutralino DM models. In order to
have the DM-induced heating rate lower than the bremsstrahlung
cooling rate at the cluster center the annihilation cross section
must be reduced by a large factor (see Colafrancesco et al. 2010
for a discussion in the case of Coma cluster).
Cored DM density profiles alleviate the over-heating problem for
neutralino DM models by reducing the electron equilibrium spectrum
$n_e(E,r)$ in the cluster central region. This makes DM models
more consistent with the cluster heating rate constraints for low
(a few to $\sim 10$ GeV) and intermediate ($\sim 40-60$ GeV)
neutralino mass, but not for high neutralino mass ($\sim 500$ GeV
or more) that still produce excess heating (see Colafrancesco et
al. 2010).

\newline{\bf Multi-Messengers}.
Stronger constraints to DM annihilation models can be set by a
multi-messenger analysis of DM signals.
Such multi-messenger constraints include, e.g., observations and
limits on positrons, antiprotons, radio and $\gamma$-rays from the
Galactic Centre region and the optical depth of CMB photons (see
discussion in Colafrancesco et al. 2010); multi-frequency (radio,
microwave, X-ray and gamma-ray) observations from satellites of
the MW, external galaxies (like M31, e.g. Borriello et al. 2009)
and cosmic background radiation, in addition to those obtained
from dwarf galaxies and galaxy clusters (see Colafrancesco 2006,
Profumo \& Ullio 2010 for reviews).\\
We stress that these multi-messenger bounds are quite robust and
not easily avoidable, and therefore they set further constraints
to any specific multi-frequency analysis of DM models, as we have
described in this review.

\newline{\bf Cosmic rays from neutralino annihilation}

Neutralino annihilation in nearby DM clumps (like e.g. the
Galactic center or galactic DM satellites) produces cosmic rays
that diffuse away and can be directly recorded by cosmic rays
experiments.\\
Some of the cosmic ray observations (at energy between 10 GeV and
a few TeV) obtained with PAMELA (Adriani et al. 2009), ATIC (Chang
et al. 2008), Fermi (Abdo et al. 2009), and HESS (Aharonian et al.
2009) show a positron excesses over background expectations
(Strong et al. 2009).\\
An attractive and widely explored possibility is that the observed
positron excesses are produced by neutralino (WIMP) DM
annihilation, even though these excesses have rather plausible
astrophysical explanations (see, e.g., Hooper et al. 2009, Yuksel
et al. 2009, Profumo 2008, Dado \& Dar 2009, Biermann et al. 2009,
Katz et al. 2009).
However, if the neutralino annihilation cross section is of the
order of thermally-averaged annihilation cross section at freeze
out, i.e. $\langle \sigma v \rangle \approx 3 \times 10^{-26}
cm^3/s$, the resulting annihilation signal is far too small to
explain the observed cosmic ray excesses.\\
A seemingly attractive solution is to postulate that DM interacts
with a light force carrier $\phi$ with fine structure constant
$\alpha_X \equiv \lambda^2/ (4 \pi)$ (see, e.g., Cirelli et al.
2009). This effectively enhances the annihilation cross section by
a quite large factor $S= {\pi \alpha_X/v_{rel} \over 1 - exp(-\pi
\alpha_X/v_{rel})}$ (usually referred to as Sommerfeld
enhancement, see e.g. Arkani-Hamed et al. 2009), where $v_{rel}$
is the DM particle relative velocity. Since the required DM mass
to explain the positron observations is $\sim$ TeV, the required
$m_{\phi}$ is $\sim$ GeV, values $S \sim 10^3$ can be obtained,
assuming $\langle \sigma v_{rel} \rangle \approx \langle \sigma v
\rangle$, that can fit the positron excess. The Sommerfeld
enhancement therefore provides an elegant mechanism for boosting
neutralino DM annihilations now.\\
Of course, for a viable solution, DM must not only annihilate with
the correct rate, but it must also be produced with the right
density and form all large-scale structure in agreement with
observations. However, Feng et al. (2009) have shown that the
required enhancement implies thermal relic densities that are too
small to be all of DM. In addition, upper bounds on possible
Sommerfeld enhancements can be derived from the observation of
elliptical galactic dark matter halos (like the case of NGC270),
and these bounds also (generically) seem to exclude enhancements
that can explain the observed positron excesses (see discussion by
Feng et al. 2009).

\subsection{Other DM scenarios}
 \label{sec:ldm}

Recent progress has greatly expanded the list of well-motivated
candidates and the possible signatures of DM (see Feng 2010 for a
recent review). Beyond any attempt to be exhaustive and/or
complete in this respect, we briefly discuss here the interesting
case of a light DM candidate, i.e. sterile neutrinos.

\begin{figure}[ht!]
 \epsfysize=8.5cm \hspace{0.0cm} \epsfbox{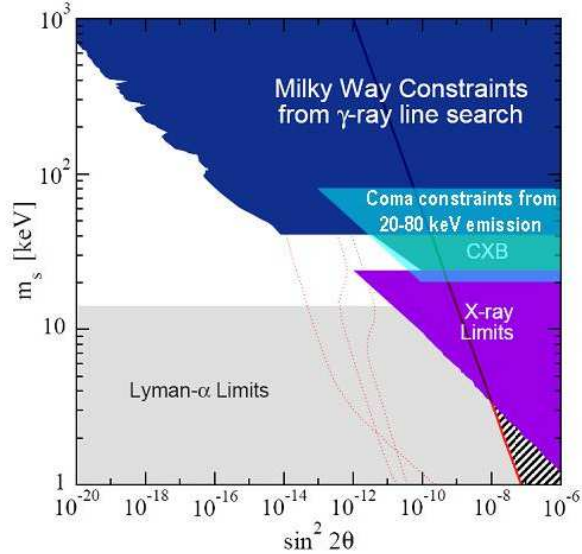}
\caption{\footnotesize{The sterile neutrino mass $m_s$ and mixing
$sin^2 (2\theta)$ parameter space, where the shaded regions are
excluded. The strongest direct bounds are shown, labeled as Milky
Way (Yuksel et al. 2008), Cosmic X-ray Background (Boyarsky et al.
2006), and X-ray limits (summarized by Watson et al. 2006). The
strongest indirect bounds (Seljak et al. 2005), Viel et al. 2006)
are shown by the grey horizontal band. The excluded
Dodelson-Widrow model (Dodelson \& Widrow 1994) is shown by the
solid line; rightward, the DM density is too high (stripes). The
dotted lines indicate example models (see Abazajian \& Fuller
2002), now truncated by the available constraints. The cyan shaded
area indicates our constraints from the HXR (soft gamma-ray) limit
on Coma (Colafrancesco 2008). Figure from Colafrancesco (2008)
adapted from the original one by Yuksel et al. (2008).
 }
 }
 \label{fig.sterile}
\end{figure}

{\bf Sterile neutrinos} may be produced in a number of ways and
their relic density depends on the sterile neutrino mass and
mixing angle, but all of the mechanisms require small masses $m_s$
and mixing angles $\theta$ for sterile neutrinos to be a viable DM
candidate.
As for the astrophysical (indirect detection) search, it is
interesting to notice that the radiative decay of sterile
neutrinos, $\nu_s \to \nu_i + \gamma$ (where $\nu_i$ indicate the
standard low-mass neutrinos), produces a narrow line emission
whose energy provides information on the sterile neutrino mass
$m_s$. Therefore, X-ray spectral observations from galaxy clusters
are a powerful tool to set contraints on sterile neutrinos in the
plane $m_s - sin^2(2\theta)$.  The available constraints on
sterile neutrinos from X-ray spectra of galaxy clusters, combined
with those obtained from the Cosmic X-ray Background, Ly$\alpha$
limits and gamma-ray line limits from the MW are shown in
Fig.\ref{fig.sterile}. The constraints from Coma observations in
the 20-80 keV band are shown by the cyan dashed area.
The possible interpretation of the intensity excess in the 8.7 keV
line (at the energy of the FeXXVI Ly-$\gamma$ line) in the
spectrum of the Galactic center observed by the Suzaku X-ray
mission in terms of decay of a sterile neutrino with mass of
$17.4$ keV and value of the mixing angle $sin^2(2 \theta) = (4.1
\pm  2.2) \cdot 10^{-12}$ (see Prokhorov \& Silk 2010), lies in
the allowed region of mass--mixing angle for DM sterile neutrino
shown is Fig.\ref{fig.sterile}.\\
This figure shows that models with lower mixing angles $\theta$
and neutrino masses $m_s$ up to a few hundreds keV or even
$\simgt$ MeV are still available. In this case, next generation
high-sensitivity hard X-ray detectors like NuSTAR and/or NeXT, or
next coming soft gamma-ray experiments
will be able to set relevant constraints to this light DM model.

\section{Dark Matter in modern cosmology: the future}

All solid evidence for DM is gravitational, and there is also
strong evidence against DM having strong or electromagnetic
interactions.
However, direct and indirect probes for DM have, so far, not yet
given a definite (positive) answer. In addition, some of the
puzzling anomalies (e.g., DAMA, PAMELA) are not easy to explain
within canonical WIMP DM models (see Feng 2009 for a discussion).
This experimental frustration and theoretical embarassement have
motivated the search for new DM candidates.
Among the logical alternatives to the WIMP paradigm (i.e., the
fact that particle physics theories designed to explain the origin
of the weak scale often naturally contain particles with the right
relic density to be DM), then, one of the most widely explored
ones is provided by hidden DM, that is, DM that has no standard
model gauge interactions (see Feng 2009, 2010 for reviews).\\
At the moment, therefore, the DM search is continuing to be a
great challenge. The direction in this effort are: direct search
experiments, indirect probes of DM signals coming from cosmic
sources, indirect indications on the nature of DM coming from
laboratory experiments, like the LHC experiment. The next future
is certainly bright in this context and expectations are rather
high.

In the cosmological and astrophysical context, the refinement for
the indirect DM search calls for a Multi$^3$ approach  in the
optimal astrophysical laboratories: i) multi-frequency to probe
the consistency of DM annihilation/decay signals across the e.m.
spectrum; ii) multi-messenger to cross-check the possible e.m.
signals in cosmic structures at different scales; iii)
multi-experiment to determine the robustness of the DM signals
with different experimental techniques.

\subsection{Multi$^3$ DM search: optimal astrophysical laboratories}
 \label{sec.optimal_labs}

The analysis of the spatial and spectral intensity of the
astrophysical signals coming from neutralino annihilation is a
powerful tool to unveil the elusive nature of DM.
However, the DM-induced signals are expected to be confused or
even overcome by other astrophysical signals originating from the
ambient gas and/or from the relativistic plasmas present in the
atmospheres of galaxy clusters and galaxies, especially when all
these components are co-spatially distributed with the DM
component.\\
An ideal system to detect DM annihilation signals would be a
system which is either devoid of diffuse emitting material (this
is the case of dark galaxies, like many dwarf spheroidals) or a
system with a clear spatial separation between the various matter
components (this is, indeed, the case of the cluster 1ES0657-556
where the spatial distribution of DM which is clearly offset with
respect to that of the intracluster gas).

A {\it multi-frequency} analysis greatly  helps to disentangle DM
signals in these optimal laboratories from signals of different
astrophysical origin.

In the {\it multi-messenger} analysis previously evoked, dark
(dwarf) galaxies are among the best sites for the astrophysical
search for the nature of DM but the relative multi-frequency SED
are usually quite dim. Definite probes of DM signals in such
systems have to be then complemented with probes coming from other
DM halos like galaxy clusters (on larger scales) and the center of
the Galaxy (on closer scales).

\begin{figure}[ht!]
 \epsfysize=7.cm \hspace{0.0cm} \epsfbox{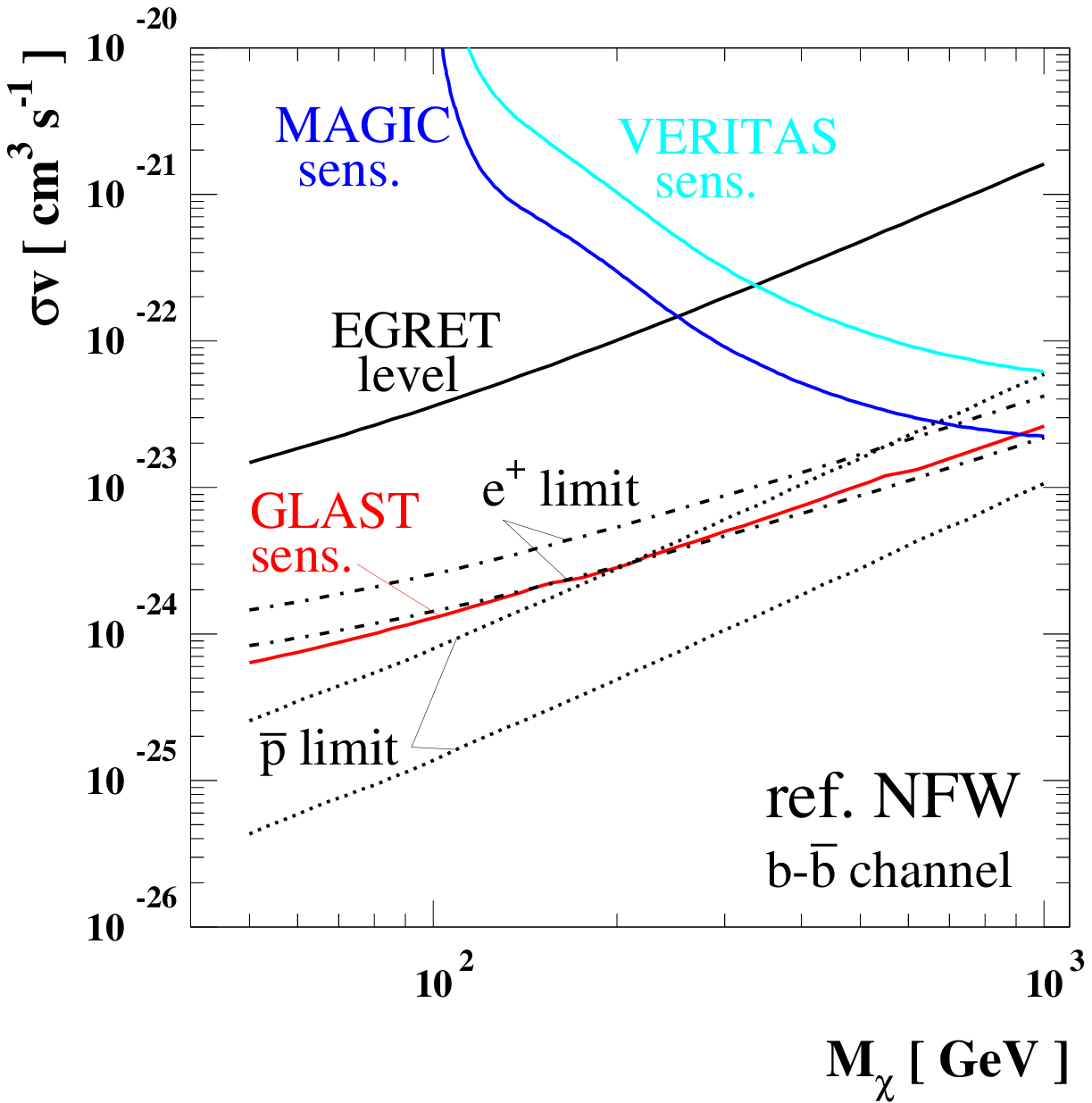}
 \epsfysize=7.cm \hspace{0.0cm} \epsfbox{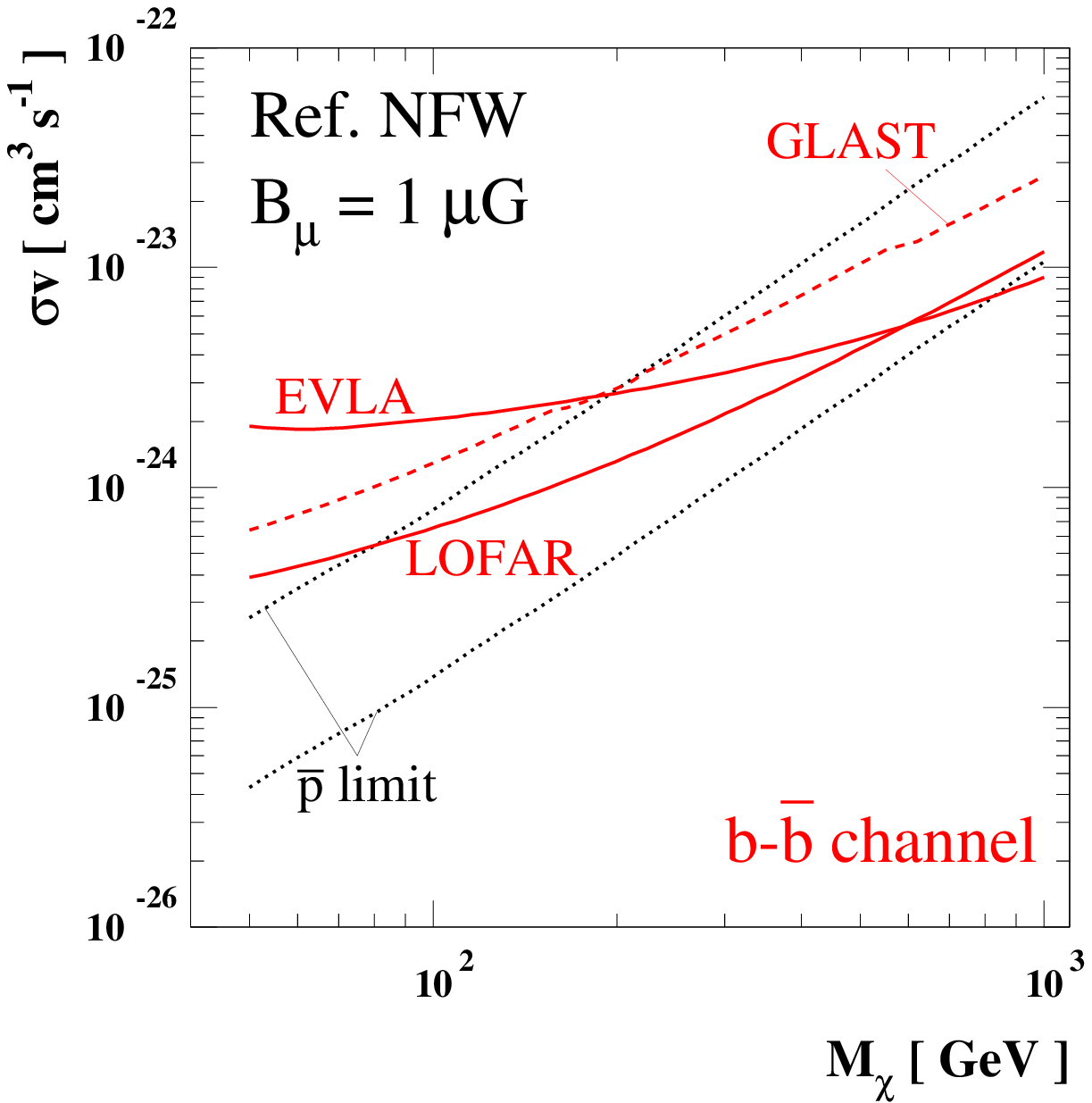}
\caption{\footnotesize{The limits in the $M_{\chi}-\langle \sigma
v \rangle$ plane set by current and future gamma-ray experiments
in a Multi$^3$ analysis of a  $b \bar{b}$ neutralino annihilation
model.
\textbf{Left}. The solid black line indicates values of $\langle
\sigma v \rangle$ required for a gamma-ray flux of the Draco dwarf
galaxy matching the 2 events in EGRET between 1 and 10~GeV, for
exposures and angular cuts as specified in the data analysis,
assuming a reference NFW halo model. The black dotted and
dot-dashed curves show the limits from the flux of, respectively,
antiprotons and positrons in the Milky Way halo (the two lines
corresponding to two different cosmic ray propagation setups); The
solid red, dark blue and light blue curves indicate the projected
sensitivity of Fermi (GLAST), MAGIC and VERITAS for Draco.
\textbf{Right}. The projected sensitivity of future diffuse radio
source searches from the direction of Draco. For comparison, we
also indicate the sensitivity of gamma-ray search experiments, and
the constraints from the antiproton and positron flux (the two
lines corresponding to two different propagation setups for the
Milky Way). (Figures and details from Colafrancesco et al. 2007).
 }
 }
 \label{fig.multiexp}
\end{figure}

The {\it multi-experiment} strategy combining radio and gamma-ray
observations of dwarf galaxies, like Draco, obtainable with high
sensitivity instruments (SKA, LOFAR, EVLA, Fermi) could set strong
constraints on the nature of the DM particles see Colafrancesco et
al. 2007 for a discussion).
Fig.\ref{fig.multiexp} shows the limits in the $M_{\chi}-\langle
\sigma v \rangle$ plane set from a Multi$^3$ analysis of
neutralino annihilation signals (taking into account the Draco
dwarf galaxy and some of the signals coming from the Milky Way)
where the most constraining frequency ranges are, in fact, the
radio and the gamma-ray bands.

As for galaxy clusters, a Multi$^3$ approach indicates that in the
case of \es  the expected gamma-ray emission associated to the DM
clumps is too low ($\simlt 1$ count vs. $\sim 10$ background
counts at $E > 1$ GeV) and cannot be resolved by Fermi from other
possible sources of gamma-ray emission, both from the cluster \es
and from AGNs (or other $\gamma$-ray emitting galaxies) in the
field.
Radio telescopes have, in principle, excellent resolution and
sensitivity to probe the different spectra and brightness
distribution of the DM-induced synchrotron emission, but the
theoretical uncertainties associated to the radio emission from
the DM clumps in \es render the interpretation of the expected
signals quite uncertain: we evaluated (Colafrancesco et al. 2007)
that the DM induced synchrotron emission from the largest DM clump
is $\sim 3-10$ mJy (for a smooth or smooth plus $50 \%$ mass
clumpiness NFW DM profile, soft DM model with $M_{\chi}=40$ GeV
with a $B=1$ $\mu$G) at $\nu = 100$ MHz, still marginally
detectable by LOFAR.
In such a context, the possible detection of the SZ$_{DM}$ effect
from this system, with the next generation high-sensitivity and
high-resolution experiments, will provide an important
complementary, and maybe unique, probe of the nature of DM.

\subsection{DM or modified gravity?}

\begin{figure}[ht!]
 \epsfysize=8.5cm \hspace{0.0cm} \epsfbox{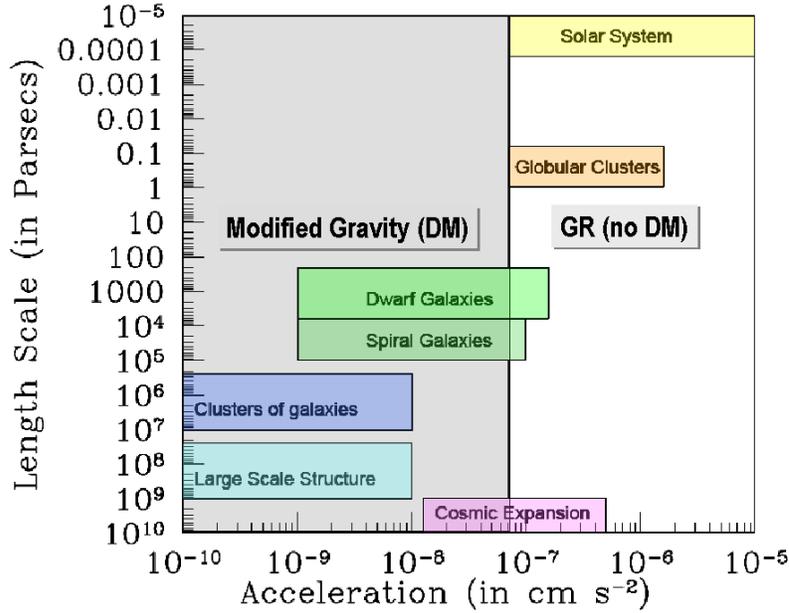}
\caption{\footnotesize{Evidence for deviations from general
relativity (GR) or evidence for the presence of Dark Matter (DM)
tend to appear in systems in which the acceleration scale is weak
(the left side, gray shaded part of the plot) at $\simlt 7 \times
10^{-8} cm s^{-2}$. There is strong evidence for either of the
above in dwarf galaxies, spiral galaxies, galaxy clusters, large
scale structures and in the expansion of the universe itself.
Figure adapted from the original one by Ferreira \& Starkman
(2009).
 }
 }
 \label{fig.modg_dm}
\end{figure}

The DM particle solution of the global DM problem is not
univocally accepted.\\
A different, more radical approach to explain the cosmological DM
problem can be taken if one notes that the evidence for missing
mass arises because of a mismatch between the gravitational field
one would predict from the observed mass distribution in the
universe and the observed gravitational field. The observed
discrepancies arise when the effective gravitational acceleration
is around, or below, the value $a_0 \sim 10^{-7} cm s^{-2}$, that
is in a regime of very weak gravitational field (see Ferreira \&
Starkman 2009 for a discussion). The point where modified gravity
scenarios are at stake, is that the Newtonian theory of gravity -
and general relativity - break down in this regime (see Fig.
\ref{fig.modg_dm}).

It has been shown that modified theory of gravity that
incorporates quantum effects can explain a crucial set of puzzling
astronomical observations (galaxy flat rotation curves, galaxy
cluster structure and the accelerated expansion of the universe),
including the "wayward" motion of the Pioneer spacecraft in our
solar system (e.g. Moffat 2005, 2006; see also Brownstein \&
Moffat 2006a,b,c).\\
Beyond the details of the specific approaches to a modified or
extended theory of gravity that have been presented so far, it is
important to notice that a relativistic theory of modified gravity
allows to make a number of specific predictions on a wide range of
scales, from the scales of compact objects such as stars or black
holes (in the limit of strong gravitational field), to the scales
of formed structures on large scales (such as galaxies and galaxy
clusters) up to the very large-scale structure scales and the
largest scale probed by CMB anisotropies (in the limit of weak
gravitational field) (see, e.g., Ferreira \& Starkman 2009,
Capozziello et al. 2009, Capozziello \& Francaviglia 2008, Schmidt
et al. 2009, among others).\\
It is, therefore, conceivable to use the next coming experimental
probes on both strong gravitational fields (in the vicinity of
very compact objects) and weak gravitational field (on large-scale
cosmic structures) to set relevant constraints on the level and on
the spatial scales at which the modification of the relativistic
theory of gravity could occur.

\section{Epilogue}

Viable DM models which are consistent with WMAP limits on
$\Omega_{DM}$ and $\Omega_{DE}$ and with the structure and
evolution of galaxy clusters are able to produce substantial
astrophysical signals especially detectable at radio, microwave,
X-ray and gamma-ray frequencies.
The constraints that the multi-frequency observations of the
optimal astrophysical laboratories for DM search can set on the
$\langle \sigma v \rangle$-$M_{\chi}$ plane, combined with the
WMAP (relic abundance) constraints on $\langle \sigma v \rangle$,
are able to efficiently restrict the available neutralino DM
models.
Additional restrictions of this plane may be obtained through a
multi-messenger approach by comparing the previous astrophysical
constraints to the constraints coming from both accelerator
physics and from other cosmological probes (e.g., the study of the
emission features from the galactic center region, galactic
satellites, dwarf galaxies, external galaxies) which are sensitive
to the amount and nature of the DM.
Direct DM detection experiments have already explored large
regions of the most optimistic SUSY models, and the planned
next-generation experiments will probably be able to explore also
the core of the SUSY models. In this context, the astrophysical
study of DM annihilation proves to be complementary, and hardly
competitive, especially when a full multi-messenger and
multi-experiment approach is chosen.\\
When combined with future accelerator results, such Multi$^3$
(multi-frequency, multi-messenger, multi-experiment) astrophysical
search might greatly help us to unveil the elusive nature of Dark
Matter.\\
The extended theory of gravity, on both an empirical and
theoretical level, seems to be the most viable alternative to a
dark cosmological model in which Dark Matter will remain still
elusive to laboratory and astrophysical probes.


\begin{theacknowledgments}
It is a pleasure to thank the colleagues met at the Gamow 2009
congress for the many interesting discussions on the structure of
a Dark universe, as well as on other related scientific problems,
that enriched my attendance to this conference.
\end{theacknowledgments}



\bibliographystyle{aipproc}   


\begin{thebibliography}{99}

\bibitem{1} Abazajian, K.N. \& Fuller, G.M. 2002, Phys. Rev. D 66, 023526
\bibitem{2} Abdo, A.A. et al. 2009, Phys. Rev. Lett. 102, 181101 (arXiv:0905.0025)
\bibitem{3} Abbott, L.F. and Sikivie, P. 1983, PhLB, 120, 133A
\bibitem{4} Adriani, O. et al. 2009, Nature 458, 607 (arXiv:0810.4995)
\bibitem{5} Aharonian, F. et al. 2009, A\&A, 508, 561
(arXiv:0905.0105)
\bibitem{6} Arkani-Hamed, N. et al. 2009, Phys. Rev. D 79, 015014
(arXiv:0810.0713)
\bibitem{7} Arnaud, M. 2005, in Proc. Enrico Fermi, International School of Physics Course CLIX,
eds. F. Melchiorri \& Y. Rephaeli (arXiv:astro-ph/0508159)
\bibitem{8} Ashman, K.M. 1992, PASP, 104, 1109
\bibitem{9} Babcock, H.W. 1939, Lick Obs. Bull., 19, 41 (No. 498)
\bibitem{10} Baltz, E. 2004, Lecture given at the 32$^{nd}$ SLAC Summer Institute (arXiv:astro-ph/0412170)
\bibitem{11} Bartelmann, M. \& Schneider, P. 2001, Phys.Rep., 340, 291
\bibitem{12} Bekenstein, J.D. 2004, Phys. Rev. D, 70 083509
\bibitem{13} Bertone, G., Zentner, A: and Silk, J. 2005, PRD, 72, 103517
\bibitem{14} Biermann, P.L. et al. 2009, Phys. Rev. Lett. 103,
061101 (arXiv:0903.4048)
\bibitem{15} Boehm, C. et al. 2004,  Phys.Rev.Lett. 92 101301
\bibitem{16} Borriello, E. et al. 2009 (arXiv:0906.2013)
\bibitem{17} Bottino, A. et al. Phys. Rev.D, 68, 043506
(arXiv:hep-ph/0304080)
\bibitem{18} Boyanovsky, D., de Vega, H.J. and Sanchez, N. 2007,
 (arXiv:0710.5180)
\bibitem{19} Boyarsky, A. et al. 2006,  MNRAS, 370, 213
\bibitem{20} Brownstein, J. R., and J. W. Moffat, 2006a, MNRAS, 367, 527-540 (arXiv:astro-ph/0507222)
\bibitem{21} Brownstein, J. R., and J. W. Moffat, 2006b, ApJ, 636,
721-741 (arXiv:astro-ph/0506370)
\bibitem{22} Brownstein, J. R., and J. W. Moffat, 2006c, Classical and Quantum Gravity 23, 3427-3436
(arXiv:gr-qc/0511026)
\bibitem{23}  Brownstein, J. R., and J. W. Moffat 2007, MNRAS, 382, 29
(arXiv:astro-ph/0702146v3)
\bibitem{24} Chang, J. et al. 2008, Nature 456, 362
\bibitem{25} Capozziello, S. \& Francaviglia, M. 2008, Gen.Rel.Grav., 40, 357
\bibitem{26} Capozziello, S. et al. 2009, MNRAS, 394, 947
\bibitem{27} Cirelli, M. et al. 2009, Nucl. Phys. B, 813, 1 (arXiv:0809.2409)
\bibitem{28} Colafrancesco, S. 2004a, in Frontiers of Cosmology,
A. Blanchard \& M. Signore Eds., p.75 and p.85
\bibitem{29} Colafrancesco, S. 2004b, A\&A, 422, L23
\bibitem{30} Colafrancesco, S. 2005, in 'Near-fields cosmology with dwarf elliptical galaxies',
IAU 198 Colloquium Proceedings, Eds.Jerjen, H.; Binggeli, B.
Cambridge: Cambridge University Press, p.229
\bibitem{31} Colafrancesco, S. 2006 ChJAS, 6a, 95C
\bibitem{32} Colafrancesco, S. 2007, What an astrophysicist can tell about the nature of Dark Matter? 2007,
to appear in Proceedings of the 6th International Workshop On The
Identification of Dark Matter (IDM 2006), 11-16 September 2006,
Rhodes, Greece
\bibitem{33} Colafrancesco, S. 2008,  ChJAS, 8, 61
\bibitem{34} Colafrancesco, S. \& Mele, B. 2001, ApJ, 562, 24
\bibitem{35} Colafrancesco, S., Marchegiani, P. \& Palladino, E. 2003, A\&A, 397, 27
\bibitem{36} Colafrancesco, S., Profumo, S. \& Ullio, P. 2006, A\&A, 455, 21
\bibitem{37} Colafrancesco, S., Profumo, S. \& Ullio, P. 2007, PRD, 75, 3513
\bibitem{38} Colafrancesco, S. et al. 2007, A\&A, 467, L1
\bibitem{39} Colafrancesco, S. et al. 2010, A\&A submitted
(arXiv:1004.1286)
\bibitem{40} Colless, M. \& Dunn, A.M. 1996, ApJ, 458, 435
\bibitem{41} Culverhouse, T., Ewans, W. and Colafrancesco, S. 2006, MNRAS, 368, 659
\bibitem{42} Dado, S. \& Dar, A. 2009, (arXiv:0903.0165)
\bibitem{43} de Bernardis, P. et al. 2000, Nature, 404, 955
\bibitem{44} Dine, M. \& Fischler, W. 1983, NuPhB, 227, 477
\bibitem{45} Dodelson, S. \& Widrow, A. 1994, Phys. Rev. Lett. 72, 17
\bibitem{46} Einasto, J. 2004, Dark Matter: early considerations
(arXiv:astro-ph/0401341)
\bibitem{47} Einasto, J. 2009, Dark Matter (arXiv:0901.0632)
\bibitem{48} Ellis, J. et al. 1983, NuPhB, 224, 427
\bibitem{49} Faber, S. M. \& Gallagher, J. S. 1979, ARA\&A, 17, 135
\bibitem{50} Feng, J. 2009, (arXiv:0908.1388)
\bibitem{51} Feng, J. et al. 2009 (arXiv:0911.0422)
\bibitem{52} Feng, J. 2019 (arXiv:1003.0904)
\bibitem{53} Ferreira, P.G. \& Starkman, G.D. 2009,
(arXiv:0911.1212)
\bibitem{54} Forman, W., Jones, C. \& Tucker, W. 1985, ApJ, 293, 102
\bibitem{55} Goldberg, H. 1983, PhLB, 131, 133
\bibitem{56} Jungman, G., Kamionkowski, M. \& Griest, K. 1996, PhR, 267, 195
\bibitem{57} Hooper, D. et al. 2009, JCAP 0901,
025 (arXiv:0810.1527)
\bibitem{58} Hu, W. \& Dodelson, S. 2002, ARA\&A, 40, 171
\bibitem{59} Kahn, F.D. \& Woltjer, L. 1959, ApJ, 130, 705
\bibitem{60} Katz, B. et al. 2009, (arXiv:0907.1686)
\bibitem{61} Komatsu, E. et al. 2010, (arXiv:1001.4538)
\bibitem{62} Kormendy, J. \& Knapp, G. R. 1987, Dark matter in the universe; Proceedings of the IAU Symposium,
Institute for Advanced Study and Princeton University, Princeton, NJ, June 24-28, 1985
\bibitem{63} Kuhn, T. S. 1970, The structure of scientific revolutions (Chicago: University
of Chicago Press, $2^{nd}$ ed., enlarged)
\bibitem{64} Milgrom, M. 1983, ApJ, 270, 365
\bibitem{65} Milgrom, M. \& Bekenstein, J. 1987, in Dark Matter in the Universe (IAU
Symposium No. 117), Eds. J. Kormendy and G.R. Knapp, (Dordrecht:
Reidel), p. 319
\bibitem{66} Moffat, J. W., 2005, JCAP, 0505, 003 (arXiv:astro-ph/0412195)
\bibitem{67} Moffat, J. W., 2006, JCAP, 0603, 004 (arXiv:gr-qc/0506021)
\bibitem{68} Munoz, C. 2004, Int.J.Mod.Phys.A, 19, 3093
(arXiv:hep-ph/0309346)
\bibitem{69} Navarro, J.F., Frenk, C. \& White, S.D.M. 1997, ApJ, 490, 493
\bibitem{70} Navarro, J.F. et al. 2004, MNRAS, 349, 1039
\bibitem{71} Oort, J. H. 1932, Bull. Astron. Inst. Netherlands, 6, 249
\bibitem{72} Oort, J.H. 1940, ApJ, 91, 273
\bibitem{73} Oort, J.H. 1960, Bull. Astr. Inst. Netherl., 15, 45
\bibitem{74} Oort, J.H. 1965, in Galactic Structure, Eds. A. Blaauw and M. Schmidt (Chicago:
Univ. Chicago Press), p. 455
\bibitem{75} Ostriker, J.P. 1993, ARA\&A, 31, 689
\bibitem{76} Ostriker, J.P. \& Peebles, P.J.E. 1973, ApJ, 186, 467
\bibitem{77} Ostriker, J.P., Peebles, P.J.E. \& Yahil, A. 1974, ApJ, 193, L 1
\bibitem{78} Ostriker, J. P. \& Steinhardt, P. 2003, New Light on Dark Matter, Science,
300, 1909
\bibitem{79} Peebles, P. J. E. 1980, The large-scale structure of the
universe (Princeton University, Princeton, N.J.)
\bibitem{80} Penzias, A. A. \& Wilson, R. W. 1965, ApJ, 142, 419
\bibitem{81} Perlmutter, S. 2000, Supernovae, Dark Energy and the Accelerating
Universe, 2000, APS. APR, G1003P
\bibitem{82} Pontecorvo, B. 1967, The Neutrino and its Importance in
Astrophysics, 1967, crps.conf...20P
\bibitem{83} Preskill, J. Wise, M.B. \& Wilczek, F, 1983, PhLB, 120, 127
\bibitem{84} Profumo, S. 2005, Phys. Rev. D, 72, 103521 (arXiv:astro-ph/0508628)
\bibitem{85} Profumo, S. 2008 (arXiv:0812.4457)
\bibitem{86} Profumo, S. \& Ullio, P. 2010 (arXiv:1001.4086)
\bibitem{87} Prokhorov, D. \& Silk, J. 2010 (arXiv:1001.0215)
\bibitem{88} Rees, M. J. 2003, Introduction, Royal Society of London Philosophical
Transactions Series A, 361, 2427
\bibitem{89} Roberts, M.S. \& Whitehurst, R.N. 1975, ApJ, 201, 327
\bibitem{90} Rubin, V.C. \& Ford, W.K. 1970, ApJ, 159, 379
\bibitem{91} Rubin, V. C., Thonnard, N., \& Ford, Jr., W. K. 1978, ApJ, 225, L107
\bibitem{92} Rubin, V. C., Ford, W. K. J., \& Thonnard, N. 1980, ApJ, 238, 471
\bibitem{93} Sanders, R. H. 1990, Astr. Astrophys. Rev., 2, 1
\bibitem{94} Sarazin, C.L. 1988, X-Ray Emissions from Clusters of
Galaxies (Cambridge Univ. Press, Cambridge)
\bibitem{95} Seljak, U. et al. 2005, Phys.Rev.D, 71, 103515  (arXiv:astro-ph/0407372)
\bibitem{96} Shaposhnikov, M. 2007 (arXiv:astro-ph/0703673)
\bibitem{97} Schmidt, F. et al. 2009 (arXiv:0908.2457)
\bibitem{98} Silk, J. 1968, Nature, 218, 453
\bibitem{99} Silk, J. 1992, Dark Populations, in IAU Symposium, Vol. 149, The
Stellar Populations of Galaxies, ed. B. Barbuy \& A. Renzini, 367
\bibitem{100} Smith, S. 1936, ApJ, 83, 23
\bibitem{101} Spergel, D. et al. 2003, ApJS, 148, 175
\bibitem{102} Srednicki, M. A. 1990, Particle physics and cosmology: dark
matter (Amsterdam: Elsevier Science Pub.)
\bibitem{103} Strong, A.W. et al. 2009 (arXiv:0907.0559)
\bibitem{104} Tinney, C.G. 1999, Nature, 397, 37
\bibitem{105} Trimble, V. 1987, ARA\&A, 25, 425
\bibitem{106} Turner, M. S. 1991, Dark matter in the universe, in
proceedings of the International Conference on Trends in
Astroparticle Physics, Santa Monica, CA, 26 Nov. - 1 Dec. 1990
\bibitem{107} Turner, M. S. 2003, Dark Matter and Dark Energy: The Critical Questions,
in Astronomical Society of the Pacific Conference Series, Vol.
291, Hubble's Science Legacy: Future Optical/Ultraviolet Astronomy
from Space, ed. K. R. Sembach, J. C. Blades, G. D. Illingworth, \&
R. C. Kennicutt, Jr., 253
\bibitem{108} van den Bergh, S. 2001, A Short History of the Missing Mass and Dark
Energy Paradigms, in Astronomical Society of the Pacific
Conference Series, Vol. 252, Historical Development of Modern
Cosmology, ed. V. J. Martinez, V. Trimble, \& M. J.
Pons-Border\'ia, 75
\bibitem{109} Viel, M. et al., Phys. Rev. Lett. 97, 071301
\bibitem{110} Yuksel, H. et al. 2008, Phys.Rev.Lett., 101, 121301 (arXiv:0706.4084)
\bibitem{111} Yuksel, H. et al. 2009, Phys. Rev. Lett. 103, 051101
(arXiv:0810.2784)
\bibitem{112} Watson et al. 2006, Phys. Rev. D 74, 033009
\bibitem{113} White, S. D. M., Frenk, C. S., Davis, M., \& Efstathiou, G. 1987,
ApJ, 313, 505
\bibitem{114} Zwicky, F. 1933, Helv.Phys.Acta, 6, 110
\bibitem{115} Zwicky, F. 1937, ApJ, 86, 217

\end{thebibliography}


\end{document}